\documentclass{article}
\usepackage[latin1]{inputenc}
\usepackage{amsmath}
\usepackage{graphicx}

\newcommand{\graphWidth}{3.45cm}
\newcommand{\mean}[1]{\left< #1 \right>}
\newcommand{\gin}{\theta^{\textrm{in}}}
\newcommand{\gout}{\theta^{\textrm{out}}}
\newcommand{\Pn}{P_{\textrm{n}}}
\newcommand{\Pm}{P_{\textrm{m}}}
\newcommand{\Pt}{P_{\textrm{t}}}
\newcommand{\ZGCC}{Z_{\textrm{GCC}}}
\newcommand{\ZGOUT}{Z_{\textrm{GOUT}}^{-}}
\newcommand{\ZGOUTr}[1]{Z_{\textrm{GOUT}}^{-(#1)}}
\newcommand{\LGCC}{L_{\textrm{GCC}}}
\newcommand{\LGOUT}{L_{\textrm{GOUT}}^{-}}

\begin{document}

\title{Probabilistic Heuristics for Disseminating Information in Networks}
\author{Alexandre~O.~Stauffer\\
Valmir~C.~Barbosa\thanks{Corresponding author ({\tt valmir@cos.ufrj.br}).}\\
\\
Universidade Federal do Rio de Janeiro\\
Programa de Engenharia de  Sistemas e Computação, COPPE\\
Caixa Postal 68511\\
21941-972 Rio de Janeiro - RJ, Brazil
}
\maketitle

\begin{abstract}
We study the problem of disseminating a piece of information through all the nodes of a network, given that it is known originally
only to a single node. In the absence of any structural knowledge on the network other than the nodes' neighborhoods,
this problem is traditionally solved by flooding all the network's edges. We analyze a recently introduced
probabilistic algorithm for flooding and give an alternative probabilistic heuristic that can lead to some cost-effective improvements, like
better trade-offs between the message and time complexities involved. We analyze the two algorithms both mathematically and
by means of simulations, always within a random-graph framework and considering relevant node-degree distributions.

\end{abstract}

\section{Introduction} \label{sec:intro}

A network can be viewed as an undirected graph $G=(N_G,E_G)$ with $n=\vert N_G \vert$ nodes and $m=\vert E_G \vert$ edges,
in which the existence of an edge $\left(u,v\right)$ represents the
possibility of bidirectional communication between nodes $u$ and $v$.
We consider the problem of disseminating a piece of information,
referred to as $I$, to all the nodes of the network, given that initially only one node, called the originator, has it.

A traditional algorithm to disseminate information in networks when nodes know their immediate neighborhoods and nothing
else is what we call uninformed flooding \cite{segall},
in allusion to the fact that its actions do in no way
take into account the structure of the network or any of its properties. In this algorithm, the originator starts by sending $I$ to its
neighbors; when receiving $I$ for the first time, each of the other nodes forwards it to its own neighbors.
This algorithm has message
complexity of $\Theta(m)$, as exactly two messages are transmitted on each edge, and worst-case time complexity of $O(n)$, the latter
related to the customary message-passing causal chains of distributed computing under full asynchronism \cite{valmir}.
Also, when edges can be assumed to have approximately equal delays associated with them, the algorithm's average waiting time, which is the average time
for a node to receive $I$, is given by the average distance from the originator in the network, that is, the average number
of edges on the shortest paths between the originator and each of the other nodes.

Our interest in this paper is to study the trade-off between the message complexity and the average waiting time when $I$ is
no longer forwarded deterministically to all of a node's neighbors, but rather is sent as the result of probabilistic decisions. Our initial
motivation has been the recent introduction of probabilistic flooding \cite{criticality}, which prescribes for each node
that it forward $I$ to each of its neighbors with fixed probability $p$. Probabilistic flooding is then also ``uninformed,'' constituting
essentially a simple stochastic generalization of the aforementioned (deterministic) uninformed flooding. We
introduce in Section~\ref{sec:heuristic} the alternative that we call heuristic flooding, which is also probabilistic in nature but,
unlike probabilistic flooding, takes the network's structure into account insofar as it can be inferred from the nodes' immediate
neighborhoods.

Naturally, flooding the network with copies of $I$ that are propagated probabilistically does not ensure that all nodes will
eventually receive a copy. There do exist applications, however, that may benefit from the aforementioned trade-off even in the
absence of such delivery guarantees. We refer the reader to \cite{criticality} and to the references therein for examples in the
area of peer-to-peer computing when $I$ is interpreted as a query and the flooding as a search.

We base our analyses of both algorithms on the generating functions of discrete
probability distributions \cite{knuth}. The formalism we utilize is the one laid down in \cite{newman}, whose
details are briefly reviewed and extended in Section~\ref{sec:preliminaries}. Our analyses are also based on what we call a flooding digraph, which
is a random directed subgraph of the network. In Section~\ref{sec:floodings}, we introduce this subgraph and use it to obtain 
our analytical results.

All our analyses are carried out in the framework of random graphs \cite{albert}. Within this framework, we give special attention
to random graphs whose node degrees are either Poisson-distributed or distributed according to a power law. The former
arise in the classic context of Erd{\H o}s and Rényi \cite{erdos1}, while the latter have recently been found to be
representative of networks like the Internet and the WWW \cite{faloutsos,medina,albert}. Simulation results are given in Section
\ref{sec:simulation} for these models of random graphs.

A comparative study of the two probabilistic approaches is given in Section~\ref{sec:comparison}, and concluding remarks
in Section~\ref{sec:conclusion}.

\section{Heuristic flooding} \label{sec:heuristic}

In probabilistic flooding, each node decides whether to send $I$ to each of its neighbors based on a fixed probability parameter
$p$. As a consequence, a node with a small degree has a smaller probability of receiving the information than a node with a large
degree. Given a node $u$ and one of its neighbors $v$, if $a$ is the degree of $u$ and $b$ the degree of $v$,
our new heuristic flooding is based on a function $h(a,b)$ representing the probability that node $u$ sends $I$ to node $v$.
Intuitively, $h(a,b)$ is expected to have a larger value when at least one of $a$ and $b$ is small than when they are both large, thus attempting
to compensate for the inherent drawback of probabilistic flooding that we just mentioned, and also to reflect the understanding that a small $a$
may signify the existence of insufficient alternative routes for $v$ to ultimately receive $I$ from $u$ even when $b$ is large.

Ideally, an accurate heuristic function would require some knowledge about the topology of the network beyond the nodes'
immediate neighborhoods; realistically, however, a node
can only be assumed to have information about its own neighbors. For this reason, heuristic flooding is defined to operate under the
assumption by each node $u$ that a neighboring node $v$ can only receive the information from $\min\left\{a,b\right\}$
of $v$'s own neighbors, for each one with equal probability $h(a,b)$, where $a$ and $b$ are the degrees of $u$ and $v$,
respectively. In other words, there are two sides to our assumption. One side is that, if $u$ decides not to forward $I$ to
$v$, then $v$ can still receive $I$ from $\min\left\{a,b\right\}-1$ of its neighbors, if any, in each case with the
same probability $h(a,b)$. The other side of the assumption is that
the other $b-\min\left\{a,b\right\}$ neighbors of $v$, if any, are unable to receive $I$ if $v$ does not send it
to them.

If we let $\alpha$ stand for the desired probability
that a node receives the information being broadcast, then, from the perspective of node $u$ upon deciding whether to send
the information to its neighbor $v$, the probability that $v$ does not receive the information (that is, $1-\alpha$) can be expressed
as
\begin{equation}
   1 - \alpha = \left[1 - \alpha + \alpha \left( 1 - h(a,b) \right)\right]^{\min\left\{a,b\right\}}.
\end{equation}
This expression indicates that $v$ does not receive the information if and only if, for each of the $\min\left\{a,b\right\}$
neighbors that the assumption says could send the information to it, either that neighbor has not itself received the information
(with probability $1 - \alpha$) or it has but decided not to forward it to $v$ (with probability $\alpha\left(1 - h(a,b)\right)$).
Hence,
\begin{equation}
   h(a,b) = \frac{1 - \left(1 - \alpha \right)^{1/\min\left\{a,b\right\}}}{\alpha}. \label{eq:heuristic}
\end{equation}

\begin{figure}[!t]
   \centering
   \includegraphics[width=8cm]{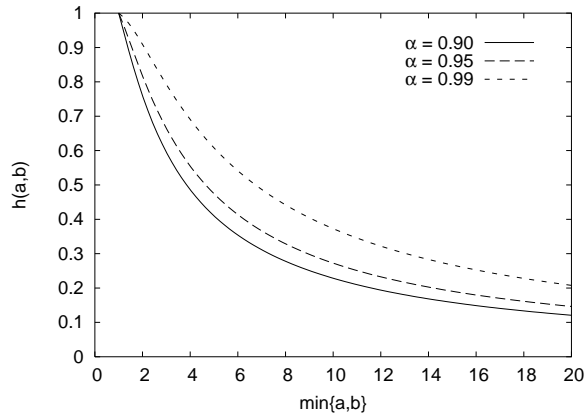}

   \caption{The heuristic function $h(a,b)$ for $\alpha=0.90,0.95,0.99$. For clarity's sake, plots are shown for a continuum of
$\min\left\{a,b\right\}$ values starting at $1$.}
   \label{fig:heuristic}
\end{figure}

This heuristic function has the property that, when both $a$ and $b$ are large, $h(a,b)$ is small, which corresponds to the
intuition that there may exist other paths along which $v$ can receive $I$, so it may not be essential that $u$ sends it to $v$.
On the other hand, when at least one of $a$ and $b$ is small, then $h(a,b)$ is large.
In particular, when at least one of $a$ and $b$ is equal to $1$, that is, $\min\left\{a,b\right\}=1$, then $h(a,b)$ is also $1$.
Illustrative plots of $h(a,b)$ against $\min\left\{a,b\right\}$ are shown in Figure~\ref{fig:heuristic} for
$\alpha=0.90,0.95,0.99$.

\section{The generating-function formalism} \label{sec:preliminaries}

Henceforth, $G$ is viewed as a random graph whose nodes all have degrees that are independent of one another and distributed identically to a
random variable $K_G$. The nodes of $G$ are assumed to be interconnected at random given their degrees, so the degrees of any two adjacent
nodes remain independent. The results from \cite{newman} reviewed in Section~\ref{sec:prelim1}, and also their extensions in
Section~\ref{sec:prelim2}, hold in the limiting case of a formally infinite number of nodes.

\subsection{Basic results} \label{sec:prelim1}

Denoting by $P_G(a)$ the probability that a randomly chosen node of $G$ has degree $a$, with $a \geq 0$, or equivalently the
probability that $K_G$ equals $a$, the generating function for the degree distribution of $G$ is $G_0(x)$ such that
\begin{equation}
   G_0(x) = \mean{x^{K_G}} = \sum_{a=0}^{n-1} x^a P_G(a), \label{eq:g0}
\end{equation}
where we employ the usual angle-bracket notation to indicate expectation.
$G_0(1)=1$ necessarily, and the average degree of $G$, denoted by $Z_G$, is the expectation
of $K_G$, that is,
\begin{equation}
   Z_G = \sum_{a=0}^{n-1} a P_G(a) = \left[\frac{d}{dx} G_0(x)\right]_{x=1} = G_0'(1). \label{eq:zg}
\end{equation}
More generally, and for $s \geq 1$, the $s$th moment of $K_G$, $\mean{K_G^s}$, is
\begin{equation}
   \mean{K_G^s} = \sum_{a=0}^{n-1} a^s P_G(a) = \left[\left(x\frac{d}{dx}\right)^s G_0(x)\right]_{x=1}. \label{eq:moment}
\end{equation}

We know from \cite{molloy1,cohen,newman} that a criterion exists according to which a random graph has almost surely a size-$\Theta(n)$ connected
component, referred to as the giant connected component (GCC). When this is the case, the other components of the graph are small if
compared to its number of nodes, that is, the fraction of nodes inside each of them is approximately $0$. The criterion in the case of $G$ is that
\begin{equation}
   \frac{\mean{K_G^2}}{Z_G} > 2. \label{eq:critical}
\end{equation}
If this inequality holds, $G$ is said to be above the phase transition. If it does not, then all the components
of $G$ are small and a GCC does not exist. In this case, $G$ is said to be below the phase transition.

Let us now consider a node at which we arrive by following a randomly chosen edge of $G$. The probability that such a node has other $b-1$
edges incident to it, with $b \geq 1$, is the expected fraction of edges incident to degree-$b$ nodes, which is given by
\begin{equation}
   \frac{b P_G(b)}{\sum_{a=0}^{n-1} a P_G(a)} = \frac{b P_G(b)}{Z_G}. \label{eq:neigh}
\end{equation}
So the number of remaining edges incident to such a node is distributed in a way that can be generated by
\begin{equation}
   G_1(x) = \sum_{b=1}^{n-1} x^{b-1} \frac{b P_G(b)}{Z_G} = \frac{G_0'(x)}{Z_G}. \label{eq:g1}
\end{equation}

Now let two nodes be called $r$-neighbors of each other, for $r \geq 1$, if the distance between them in $G$ is $r$. The case of
$r=1$ is simply the case of neighbors in $G$. Given a neighbor of a randomly chosen node $u$, the expected number of that neighbor's
other neighbors (i.e., excluding $u$) is
\begin{equation}
   \sum_{b=1}^{n-1} (b-1) \frac{b P_G(b)}{Z_G} = \left[\frac{d}{dx} G_1(x)\right]_{x=1}=\frac{G_0''(1)}{Z_G},
\end{equation}
so the expected number of $2$-neighbors of $u$, which we denote by $Z_G^{(2)}$, is given by
\begin{eqnarray}
   Z_G^{(2)} & = & \sum_{a=0}^{n-1}aP_G(a)\sum_{b=1}^{n-1} (b-1) \frac{b P_G(b)}{Z_G} \nonumber \\
             & = & Z_G \sum_{b=1}^{n-1} (b-1) \frac{b P_G(b)}{Z_G}. \label{eq:zg2}
\end{eqnarray}
In general, we denote a node's expected number of $r$-neighbors by $Z_G^{(r)}$. Clearly, $Z_G^{(1)}=Z_G$, and (\ref{eq:zg2}) can be
generalized to yield
\begin{eqnarray}
   Z_G^{(r)} & = & Z_G^{(r-1)} \sum_{b=1}^{n-1} (b-1) \frac{b P_G(b)}{Z_G} \nonumber \\
             & = & Z_G \left( \sum_{b=1}^{n-1} (b-1) \frac{b P_G(b)}{Z_G}\right)^{r-1} \nonumber \\
             & = & Z_G \left( \frac{Z_G^{(2)}}{Z_G}\right)^{r-1}. \label{eq:zr}
\end{eqnarray}
Note that here we have repeatedly taken into account, as we consider nodes' neighbors that are progressively farther from the initial,
randomly chosen node $u$, that the probability that any two nodes involved in the process are in fact the same node is $0$ in the limit 
as $n \to \infty$.

We can then obtain an approximation for the expected path length of $G$ by summing the
values of $Z_G^{(r)}$ from $r=1$ up until the sum becomes equal to $n-1$. When this happens, the current value of $r$ can be taken as
$L_G$, the desired expected path length. Thus,
\begin{equation}
   \sum_{r=1}^{L_G} Z_G^{(r)} = n-1,
\end{equation}
which yields
\begin{equation}
   L_G = \frac{\ln\left[\left(\frac{n-1}{Z_G}\right)\left(\frac{Z_G^{(2)}}{Z_G} - 1\right) + 1\right]}{\ln(Z_G^{(2)}/Z_G)}.
   \label{eq:lg}
\end{equation}

We henceforth assume that $G$ is above the phase transition, that is, $G$ almost surely has a GCC.
Given two adjacent nodes $u$ and $v$, let the reach of $u$ through $v$ be the set
of nodes reachable by a path starting at $u$ whose first edge is $(u,v)$. A randomly
chosen node is outside the GCC if and only if all of its neighbors are also outside the GCC, that is, it has
a small, size-$o(n)$ reach through each of its neighbors. Denoting by
$q$ the probability that this happens for each of those neighbors, we can express $q$ as
\begin{equation}
   q = \sum_{a=1}^{n-1} q^{a-1} \frac{a P_G(a)}{Z_G} = G_1(q), \label{eq:gccq}
\end{equation}
that is, a node has a small reach through one of its neighbors if and only if that neighbor itself has a small reach through
each of its other neighbors. As a result, the probability that a randomly chosen node is outside the GCC is
\begin{equation}
   \sum_{a=0}^{n-1} q^a P_G(a) = G_0(q), \label{eq:gcc1}
\end{equation}
and the fraction of nodes inside the GCC of $G$, denoted by $\theta_G$, is given by
\begin{equation}
   \theta_G = 1 - G_0(q). \label{eq:gcc}
\end{equation}

\subsection{Extensions within the GCC} \label{sec:prelim2}

One further assumption that we make in this paper is that the originator, the node that initially has the information and
starts the dissemination, is inside the GCC. In this case our analyses must be based on a degree distribution that is conditioned
upon this membership in the GCC. The conditional probability that a node inside the
GCC has degree $a$, denoted by $P_G(a \mid \textrm{GCC})$, can be written using Bayes' rule as
\begin{equation}
   P_G(a \mid \textrm{GCC}) = \frac{P_G(\textrm{GCC} \mid a) P_G(a)}{P_G(\textrm{GCC})}.
\end{equation}
Here $P_G(\textrm{GCC})$ is the probability that a randomly chosen node is inside the GCC, that is,
$\theta_G$. $P_G(\textrm{GCC} \mid a)$, in turn, is the probability that a degree-$a$ node is inside the GCC.
Such a node is outside the GCC if and
only if it has a small reach through each of its neighbors, which occurs with probability $q$ for each neighbor. Then a degree-$a$
node is inside the GCC with probability $P_G(\textrm{GCC} \mid a ) = 1 - q^a$. We then have
\begin{equation}
   P_G(a \mid \textrm{GCC}) = \frac{\left(1-q^a\right) P_G(a)}{\theta_G}.
\end{equation}

We can now obtain the generating function for the degree of a randomly chosen node inside the GCC, which we denote by
$G_0^{\textrm{GCC}}(x)$, as
\begin{eqnarray}
   G_0^{\textrm{GCC}}(x) & = & \sum_{a=0}^{n-1} x^a \frac{\left(1-q^a\right) P_G(a)}{\theta_G} \nonumber \\
                         & = & \frac{G_0(x) - G_0(qx)}{\theta_G}.
\end{eqnarray}
Also, following our earlier steps, this can be used to calculate the expected number of neighbors and $2$-neighbors of a randomly chosen
node inside the GCC, which we refer to as $\ZGCC$ and $\ZGCC^{(2)}$, respectively. We obtain
\begin{equation}
   \ZGCC = \left[\frac{d}{dx} G_0^{\textrm{GCC}}(x)\right]_{x=1} = \frac{Z_G - q G'_0(q)}{\theta_G}
\end{equation}
and
\begin{equation}
   \ZGCC^{(2)} = \ZGCC \sum_{a=1}^{n-1} (a-1) \frac{a P_G(a)}{Z_G}.
   \label{eq:z2gcc}
\end{equation}
We can also retrace the steps that led us to (\ref{eq:lg}), and obtain $\LGCC$, the expected path length inside the GCC:
\begin{equation}
   \LGCC = \frac{\ln\left[\left(\frac{n\theta_G-1}{\ZGCC}\right)\left(\frac{\ZGCC^{(2)}}{\ZGCC} - 1\right) + 1\right]}{\ln(\ZGCC^{(2)}/\ZGCC)}.
   \label{eq:lgcc}
\end{equation}

\section{Mathematical analysis} \label{sec:floodings}

We start by considering a generic probabilistic algorithm for disseminating a piece of information $I$ on the network represented by
$G$. In this algorithm, when a node of degree $a$
receives $I$ for the first time, it forwards $I$ to each of its neighbors with probability $f(a,b)$, where $b$ is the degree of
the prospective recipient. This process induces the appearance of a random directed subgraph $F=(N_F,E_F)$ of $G$ that we call the
flooding digraph. This digraph has the same nodes as $G$ (that is, $V_F=V_G$) and its edges are such that, for $(u,v) \in E_G$, the
directed edge $(u \to v)$ exists in $E_F$ with probability $f(a,b)$, given that $a$ and $b$ are the degrees of nodes
$u$ and $v$ in $G$, respectively.

Clearly, this generic algorithm can stand for both probabilistic flooding and heuristic flooding. In the former case, $f(a,b)=p$
regardless of $a$ or $b$; in the latter, $f(a,b)$ is the heuristic function $h(a,b)$ given
by (\ref{eq:heuristic}). As probabilistic flooding is a special case of heuristic flooding, for the remainder of
this section we concentrate solely on the latter and use the $h(a,b)$ of (\ref{eq:heuristic}) instead of $f(a,b)$.

\subsection{Random digraphs}
Let us first review some of the properties of random digraphs in general and also how they apply to the case of $F$.

For $u$ a node of a digraph, its in-neighbors are those nodes from
which an edge exists directed toward $u$; its out-neighbors are those nodes toward which an edge exists directed from $u$. Moreover, a node
is an $r$-in-neighbor of $u$ if the directed distance from it to $u$ is $r$. Similarly, a node is an $r$-out-neighbor of
$u$ if the directed distance from $u$ to it is $r$.
When a directed path exists from node $u$ to node $v$, we say that $v$ is reachable from $u$ (equivalently, $u$ reaches $v$).

A connected component of the undirected graph that underlies a digraph (i.e., the graph that we obtain when edge directions are ignored)
is called a weakly connected component of the digraph. Thus a digraph has a size-$\Theta(n)$ weakly connected component,
referred to as the giant weakly connected component (GWCC), if and only if the undirected graph that underlies it has a GCC.

\begin{figure}[!t]
   \centering
   \includegraphics[width=8cm]{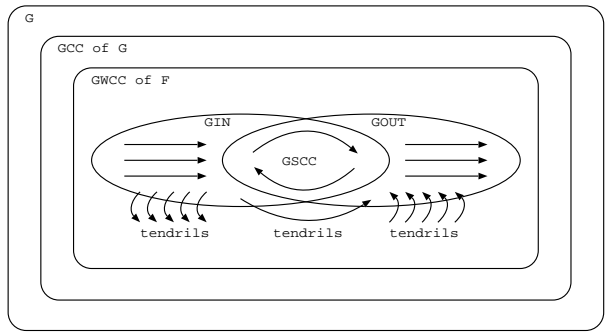}

   \caption{The structure of the GWCC of $F$ and its relation to $G$ and to the GCC of $G$.}
   \label{fig:digraphDiag}
\end{figure}

The GWCC of a digraph has four distinguished types of sub-digraphs. In the case of $F$, they are as illustrated in Figure~\ref{fig:digraphDiag}
and such that:
\begin{itemize}
   \item The giant strongly connected component (GSCC) is the largest sub-digraph of the GWCC that is maximal with respect to the property
         that any of its nodes is reachable from any other.
   \item The giant in-component (GIN), comprising a fraction $\gin_F$ of the nodes of $G$ (or a fraction $\gin_F/\theta_G$ of
         the nodes of the GCC of $G$), contains all the nodes that can reach the GSCC (including, by definition, those of the GSCC).
   \item The giant out-component (GOUT), comprising a fraction $\gout_F$ of the nodes of $G$ (or a fraction $\gout_F/\theta_G$ of
         the nodes of the GCC of $G$), contains all the nodes that are reachable from the GSCC (including, by definition, those of the GSCC).
   \item Each of the so-called tendrils consists of some of the remaining nodes \cite{dorogovtsev}.
\end{itemize}

\subsection{Number of nodes reached} \label{sec:pn}

We now consider the expected fraction of nodes of $G$ that receive $I$ in heuristic flooding. Recall that, by assumption, $G$ almost
surely has a GCC and the originator is one of the nodes of the GCC.
Two cases must be considered.
The first case corresponds to the originator being outside the GIN, belonging therefore either to
the portion of the GOUT that does not intersect the GSCC or to a tendril. This case happens with probability $1-\gin_F/\theta_G$
and leads to a negligibly small number of nodes reached by the flooding.
The second case is the case in which the originator is
inside the GIN. It occurs with probability $\gin_F/\theta_G$ and the flooding necessarily reaches
all the nodes of the GOUT. Nodes outside the GOUT may also be reached but contribute negligibly, since they belong either to
the portion of the GIN that does not intersect the GSCC or to a tendril. So, in essence, the probability that a node
is reached, denoted by $\Pn$, is
\begin{equation}
   \Pn = \frac{\gin_F \gout_F}{\theta_G^2}. \label{eq:pn}
\end{equation}

In order to calculate $\gin_F$ and $\gout_F$, let us first define the ancestry of a node as the set of nodes from which it can
be reached in $F$, and the descent of a node as the set of nodes reachable from it in $F$. Then a randomly chosen node is inside the
GIN if and only if it has a large, size-$\Theta(n)$ descent. Now consider a randomly chosen node $u$ having degree $a$ in $G$. We say
that a neighbor $v$ of $u$ in $G$ is a dead-end with respect to $u$ if either $(u \to v)$ is not an edge of $F$ or
it is but $v$ has a small descent. If $b$ is the degree of $v$ in $G$, we denote
by $q_b^{\textrm{out}}$ the conditional
probability that $v$ has a small descent, given that it has a directed edge in $F$ incoming from $u$.
The probability that a degree-$b$ node is a dead-end with respect to a degree-$a$ neighbor in $G$, which we denote by
$w_{a,b}^{\textrm{out}}$, is then
\begin{equation}
   w_{a,b}^{\textrm{out}} = 1 - h(a,b) + h(a,b) q_b^{\textrm{out}}. \label{eq:wout}
\end{equation}
But the probability that a node's neighbor in $G$ has degree $b$ is given by (\ref{eq:neigh}), so the probability that a given neighbor of
a degree-$a$ node is a dead-end is
\begin{equation}
   \sum_{b=1}^{n-1}w_{a,b}^{\textrm{out}}\frac{b P_G(b)}{Z_G}. \label{eq:badoutneigh}
\end{equation}
We also know that such a degree-$b$ node has a small descent if and only if each of its $b-1$ other neighbors in $G$
is itself a dead-end with respect to it. Thus, $q^{\textrm{out}}_b$ is such that
\begin{equation}
   q^{\textrm{out}}_b = \left(\sum_{c=1}^{n-1}w_{b,c}^{\textrm{out}}\frac{c P_G(c)}{Z_G}\right)^{b-1}. \label{eq:ginq}
\end{equation}

It now suffices to recognize that a randomly chosen node is outside the GIN if and only if all of its neighbors in $G$ are dead-ends
with respect to it. The fraction of nodes inside the GIN is then given by
\begin{equation}
   \gin_F = 1 - \sum_{a=0}^{n-1} \left(\sum_{b=1}^{n-1}w_{a,b}^{\textrm{out}}\frac{b P_G(b)}{Z_G} \right)^a P_G(a). \label{eq:gin}
\end{equation}
The equations in (\ref{eq:wout}) and (\ref{eq:ginq}) lead to a nonlinear system of $n-1$ equations (letting $b=1, \ldots, n-1$ in (\ref{eq:ginq}))
on $n-1$ variables ($q_1^{\textrm{out}}$ through $q_{n-1}^{\textrm{out}}$). A solution of this system within $[0,1]^{n-1}$ can be used, via (\ref{eq:wout}),
to calculate $\gin_F$ in (\ref{eq:gin}).

The calculation of $\gout_F$ follows a completely analogous development, since a randomly chosen node is inside the GOUT if and only if it
has a large, size-$\Theta(n)$ ancestry. This leads to
\begin{equation}
   w_{a,b}^{\textrm{in}} = 1 - h(b,a) + h(b,a) q_b^{\textrm{in}}, \label{eq:win}
\end{equation}
where
\begin{equation}
   q^{\textrm{in}}_b = \left(\sum_{c=1}^{n-1} w_{b,c}^{\textrm{in}} \frac{c P_G(c)}{Z_G}\right)^{b-1}
\end{equation}
is the conditional probability that a degree-$b$ node having a degree-$a$ neighbor in $G$ has a small ancestry,
given that in $F$ it has an edge outgoing to that degree-$a$ neighbor.
The counterpart of (\ref{eq:gin}) is then
\begin{equation}
   \gout_F = 1 - \sum_{a=0}^{n-1} \left(\sum_{b=1}^{n-1} w_{a,b}^{\textrm{in}} \frac{b P_G(b)}{Z_G} \right)^a P_G(a).
\end{equation}

\subsection{Number of messages sent} \label{sec:pm}

Let $\Pm$ be the ratio of the expected number of messages sent by heuristic flooding
to the expected number of messages sent by uninformed flooding. Given that the originator of both algorithms
is inside the GCC, the expected number of messages sent by uninformed flooding is $\ZGCC n \theta_G$. Letting $Z_{\textrm{r}}$ be the
expected out-degree in $F$ of the nodes reached by heuristic flooding, the expected number of messages in this case is
$Z_{\textrm{r}} n \theta_G \Pn$. We then have
\begin{equation}
   \Pm = \frac{Z_{\textrm{r}} \Pn}{\ZGCC}.
\end{equation}

The value of $Z_{\textrm{r}}$ can be approximated by the average out-degree of the nodes inside the GOUT, which we denote by $\ZGOUT$.
Letting $P_F^{-}(i\mid\textrm{GOUT})$ be the conditional probability that a
randomly chosen node has out-degree $i$ in $F$, given that it is inside the GOUT, $\ZGOUT$ can be expressed as
\begin{equation}
   \ZGOUT = \sum_{i=0}^{n-1} i P_F^{-}(i\mid\textrm{GOUT}). \label{eq:zgout}
\end{equation}
If we let $P_F^{-}(i\mid a,\textrm{GOUT})$ be the conditional probability that a node has out-degree $i$ in $F$, given that it has degree $a$ in
$G$ and that it is inside the GOUT, then $P_F^{-}(i\mid\textrm{GOUT})$ can be written as
\begin{equation}
   P_F^{-}(i\mid\textrm{GOUT}) = \sum_{a=i}^{n-1} P_F^{-}(i\mid a,\textrm{GOUT}) P_G(a\mid\textrm{GOUT}).
\end{equation}

Now recall that every node's degree in $G$ is by assumption independent and identically distributed with respect to all others'.
Thus, letting $h_a^-$ be the expectation of the heuristic function $h(a,b)$ as $b$ varies, that is,
\begin{equation}
   h_a^- = \sum_{b=0}^{n-1} h(a,b) \frac{b P_G(b)}{Z_G},
\end{equation}
the conditional probability that a degree-$a$ node has out-degree $i$ in $F$, which we denote by $P_F^-(i \mid a)$, is given by the
binomial distribution:
\begin{equation}
   P_F^-(i \mid a) = \binom{a}{i} \left(h_a^-\right)^i \left(1-h_a^-\right)^{a-i}.  \label{eq:aux1}
\end{equation}
But given a node inside the GOUT, (\ref{eq:neigh}) still gives the probability that one of its neighbor's degree is $b$ in $G$.
Therefore, a node with degree $a$ in $G$ that is inside the GOUT has out-degree $i$ in $F$ with probability given
by (\ref{eq:aux1}), that is, $P_F^{-}(i \mid a,\textrm{GOUT}) = P_F^{-}(i \mid a)$, which leads to
\begin{equation}
   P_F^{-}(i \mid \textrm{GOUT}) = \sum_{a=i}^{n-1} P_F^{-}(i \mid a) P_G(a \mid \textrm{GOUT}). \label{eq:aux2}
\end{equation}

We can then re-write (\ref{eq:zgout}) as
\begin{eqnarray}
   \ZGOUT & = & \sum_{i=0}^{n-1} i \sum_{a=i}^{n-1} \binom{a}{i} \left(h_a^-\right)^i \left(1-h_a^-\right)^{a-i} P_G(a \mid \textrm{GOUT}) \nonumber \\
          & = & \sum_{a=0}^{n-1} P_G(a \mid \textrm{GOUT}) \sum_{i=0}^{a} i \binom{a}{i} \left(h_a^-\right)^i \left(1-h_a^-\right)^{a-i},
\end{eqnarray}
where the rightmost summation gives the expected out-degree in $F$ of a node whose degree in $G$ is $a$, i.e., $ah_a^-$. Therefore,
\begin{equation}
   \ZGOUT = \sum_{a=0}^{n-1} P_G(a \mid \textrm{GOUT}) a h_a^-. \label{eq:zgout2}
\end{equation}

Now, if we apply Bayes' rule and write $P_G(a \mid \textrm{GOUT})$ as
\begin{eqnarray}
   P_G(a \mid \textrm{GOUT}) & = & \frac{P_G(\textrm{GOUT} \mid a) P_G(a)}{P_G(\textrm{GOUT})} \nonumber \\
                             & = & \frac{P_G(\textrm{GOUT} \mid a) P_G(a)}{\theta_F^{\textrm{out}}}, \label{eq:aux3}
\end{eqnarray}
and furthermore recognize that a node of degree $a$ in $G$ is outside the GOUT (with probability $1-P_G(\textrm{GOUT} \mid a)$) if and only if each
of its neighbors in $G$ is like the degree-$b$ nodes at the end of Section~\ref{sec:pn}, and also that this occurs for each such node with
probability
\begin{equation}
   \sum_{b=1}^{n-1}w_{a,b}^{\textrm{in}}\frac{b P_G(b)}{Z_G},
\end{equation}
then we obtain
\begin{equation}
   P_G(\textrm{GOUT} \mid a) = 1 - \left(\sum_{b=1}^{n-1}w_{a,b}^{\textrm{in}}\frac{b P_G(b)}{Z_G}\right)^a,
   \label{eq:aux4}
\end{equation}
culminating in
\begin{equation}
   \ZGOUT = \sum_{a=0}^{n-1} a h_a^- \left[ 1 - \left(\sum_{b=1}^{n-1}w_{a,b}^{\textrm{in}}\frac{b P_G(b)}{Z_G}\right)^a \right]
            \frac{P_G(a)}{\theta_F^{\textrm{out}}}.
            \label{eq:zgoutfinal}
\end{equation}

\subsection{Average waiting time} \label{sec:pt}
As we mentioned in Section~\ref{sec:intro}, assessing a distributed algorithm's time-related complexities
in a fully general asynchronous setting requires that message-passing
causal chains be considered \cite{valmir}. In our current context, and taking into account our current knowledge of random graphs and
their analysis, obtaining accurate analytical estimates of those complexities for flooding seems infeasible. Our choice is then to
settle for a less general form of asynchronism in which the delay associated with the delivery of a message over an edge of $G$ is roughly
constant for all of $G$'s edges. In this case, the waiting time before a node is reached by a flooding can be taken to be proportional
to the distance in $G$ from the originator to that node.

In order to compare the expected waiting time of heuristic flooding with that of uninformed flooding, we calculate the
ratio $\Pt$ of the expected path length from the originator in $F$ to the expected path length from the originator in $G$.
Recall that, even though we employ the common denomination as a path, in the case of $F$ paths are directed, while in $G$ they
are undirected.

For uninformed flooding with originator inside the GCC, the expected path length from the originator can be approximated by
$\LGCC$ as in (\ref{eq:lgcc}). In order to obtain the corresponding expression for heuristic flooding,
we proceed as we did for $\Pn$ and $\Pm$ and consider only the case in which the
originator is inside the GIN, which occurs with probability $\theta_F^{\textrm{in}}/\theta_G$. We also assume that
only the nodes inside the GOUT can be reached by the flooding. Letting $\LGOUT$ be the expected path length from the
originator under these assumptions, $\Pt$ is such that
\begin{equation}
   \Pt = \frac{\theta_F^{\textrm{in}}\LGOUT}{\theta_G\LGCC}.
   \label{eq:pt}
\end{equation}

The average out-degree inside the GOUT, $\ZGOUT$, is given by (\ref{eq:zgoutfinal}). In order to obtain the
expected number of $2$-out-neighbors of a randomly chosen node inside the GOUT, denoted by
$\ZGOUTr{2}$, we first consider two adjacent nodes $u_1$ and $u_2$ having degrees $a_1$ and $a_2$, respectively, in $G$.
If the directed edge $(u_1 \to u_2)$ does not exist
in $F$, which occurs with probability $1-h(a_1,a_2)$, then $u_1$ has no $2$-out-neighbors reachable through $u_2$. On the other hand,
if the directed edge does exist, and this has probability $h(a_1,a_2)$ of occurring, then the number of $2$-out-neighbors that $u_1$ can reach through
$u_2$ is expected to be $\left(a_2-1\right)h_{a_2}^-$. So each neighbor of $u_1$ in $G$ provides an expected number of
$2$-out-neighbors to it that is given by
\begin{equation}
   \sum_{a_2=1}^{n-1} \left(a_2-1\right) h_{a_2}^- h(a_1,a_2)\frac{a_2 P_G(a_2)}{Z_G}.
\end{equation}
Now let $t_{a_1,a_2}(x)$ be the function
\begin{equation}
   t_{a_1,a_2}(x) = x \left(a_2-1\right) h_{a_2}^- h(a_1,a_2) \frac{a_2 P_G(a_2)}{Z_G}.
\end{equation}
Then $\ZGOUTr{2}$ is given by
\begin{equation}
   \ZGOUTr{2} = \sum_{a_1=0}^{n-1} a_1 P_G(a_1 \mid \textrm{GOUT}) \sum_{a_2=1}^{n-1} t_{a_1,a_2}(1).
   \label{eq:z2gout}
\end{equation}

Proceeding likewise, we see that, for $r>1$, the expected number of $r$-out-neighbors of a node inside the GOUT is given by
\begin{eqnarray}
   \lefteqn{\ZGOUTr{r} = \sum_{a_1=0}^{n-1} a_1 P_G(a_1 \mid \textrm{GOUT})} \hspace{1.0in} \nonumber \\
   && \sum_{a_2=1}^{n-1} t_{a_1,a_2}\left(\sum_{a_3=1}^{n-1} t_{a_2,a_3}\left(\cdots\sum_{a_r=1}^{n-1} t_{a_{r-1},a_r}(1)\right)\right).
   \label{eq:zrgout}
\end{eqnarray}
But unlike the $Z_G^{(r)}$ of (\ref{eq:zr}) or the $\ZGCC^{(r)}$ that implicitly led to (\ref{eq:lgcc}), now the sequence
$\ZGOUT, \ZGOUTr{2}, \ldots$ is not a geometric progression
and $\LGOUT$ cannot be expressed by an equation analogous to (\ref{eq:lgcc}). We could, however, in principle, use
(\ref{eq:zrgout}) to obtain $\lfloor \LGOUT \rfloor$ and $\lceil \LGOUT \rceil$ by extending the sum
$\ZGOUT + \ZGOUTr{2} + \cdots$ up until a number greater than $n-1$ were obtained. The last two values
to go into the sum would be $\ZGOUTr{\lfloor L_{\textrm{GOUT}} \rfloor}$ and
$\ZGOUTr{\lceil L_{\textrm{GOUT}} \rceil}$. But the time complexity of calculating
$\ZGOUTr{r}$ is $O(n^r)$, so this method is not really practical.

A way out does exist, however, in some special cases. In the case of probabilistic flooding, for example, we have $h(a_1,a_2)=p$ for
$1 \leq a_1, a_2 \leq n-1$, so $t_{a_1,a_2}(x)$ is independent of $a_1$, leading (\ref{eq:zrgout}) to be simplified as
\begin{equation}
   \ZGOUTr{r} = \sum_{a_1=0}^{n-1} a_1 P_G(a_1 \mid \textrm{GOUT}) \left( \sum_{a_2=1}^{n-1} t_{a_1,a_2}(1)\right)^{r-1}.
\end{equation}
In this case, the sequence $\ZGOUTr{1}, \ZGOUTr{2}, \ldots$ is indeed a geometric progression, so we
can proceed as we did for the undirected case and obtain
\begin{equation}
   \LGOUT = \frac{\ln\left[\left(\frac{n\gout_F-1}{\ZGOUT}\right)\left(\frac{\ZGOUTr{2}}{\ZGOUT} - 1\right) + 1\right]}{\ln(\ZGOUTr{2}/\ZGOUT)}.
   \label{eq:lgout2}
\end{equation}

The case of a Poisson degree distribution is also amenable to further analysis, since we can approximate $\ZGOUTr{r}$ by
$\ZGOUT \rho^{r-1}$, where $\rho$ is
the expected number of out-neighbors that a node chosen by moving along the direction of
a randomly chosen edge of $F$ has \cite{karp}. Let us consider such an edge of $F$. The joint probability that the node from which
this edge outgoes has degree $a_1$ in $G$ and that the node to which it incomes has degree $a_2$ in $G$ is denoted
by $P_G(a_1,a_2 \mid  e)$, where $e$ represents the event that the edge exists in $F$. Using
Bayes' rule, this probability can be written as
\begin{eqnarray}
   P_G(a_1,a_2 \mid  e) & = &  \frac{P_G(e \mid a_1,a_2)P_G(a_1,a_2)}{P(e)} \nonumber \\
                        & = &  \frac{h(a_1,a_2) a_1 P_G(a_1) a_2 P_G(a_2)}{Z_G \sum_{b=1}^{n-1} h_b^- b P_G(b)},
\end{eqnarray}
which leads to
\begin{equation}
   \rho = \sum_{a_1=1}^{n-1} \sum_{a_2=1}^{n-1} (a_2-1)h_{a_2}^-\frac{h(a_1,a_2) a_1 P_G(a_1) a_2 P_G(a_2)}
                                                                     {Z_G \sum_{b=1}^{n-1} h_b^- b P_G(b)},
\end{equation}
and finally to
\begin{equation}
   \LGOUT = \frac{\ln\left[\left(\frac{n\gout_F-1}{\ZGOUT}\right)\left(\rho - 1\right) + 1\right]}{\ln \rho}.
   \label{eq:lgout1}
\end{equation}

\section{Simulation results}  \label{sec:simulation}

In our simulations we have used random graphs as models for networks \cite{albert}. The network may thus not be
connected, so simulations have only been carried out inside the largest connected component of the random graph.
Also, in order to ensure that such a component encompasses a large number of nodes, we have restricted ourselves
to graphs for which a GCC is almost surely guaranteed to exist. Our interest has been to analyze $\Pn$, $\Pm$, and $\Pt$, and to this end
we have simulated probabilistic flooding for $p=0.30$, $0.60$, $0.90$ and heuristic flooding for
$\alpha=0.90$, $0.95$, $0.99$.

We have considered two random-graph models. In the first model $G$ is generated on $n$ nodes by creating an edge between
any two distinct nodes with fixed probability $z/(n-1)$ for suitable $z$. A node in the resulting graph has degree $a$ with the Poisson
probability of mean $z$ \cite{bollobas}, that is, $P_G(a)=e^{-z}z^a/a!$. Also, we have $Z_G=z$ and $\mean{K_G^2}=z^2 + z$,
which by (\ref{eq:critical}) implies that we need $z > 1$ for the GCC to almost surely exist.

\begin{figure*}[!t]
   \centering
   \begin{tabular}{ccc}
   \resizebox*{\graphWidth}{!}{\includegraphics{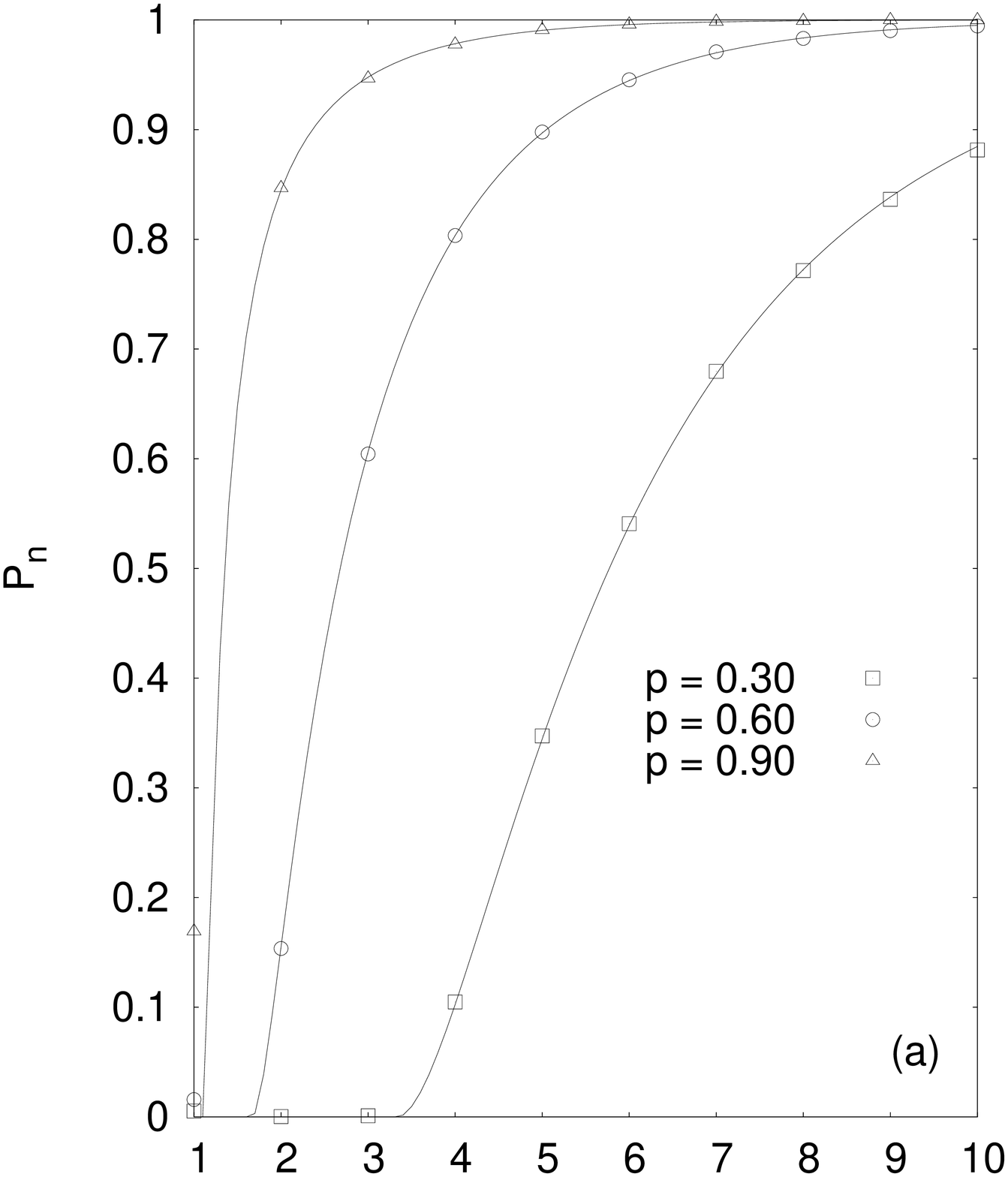}} &
   \resizebox*{\graphWidth}{!}{\includegraphics{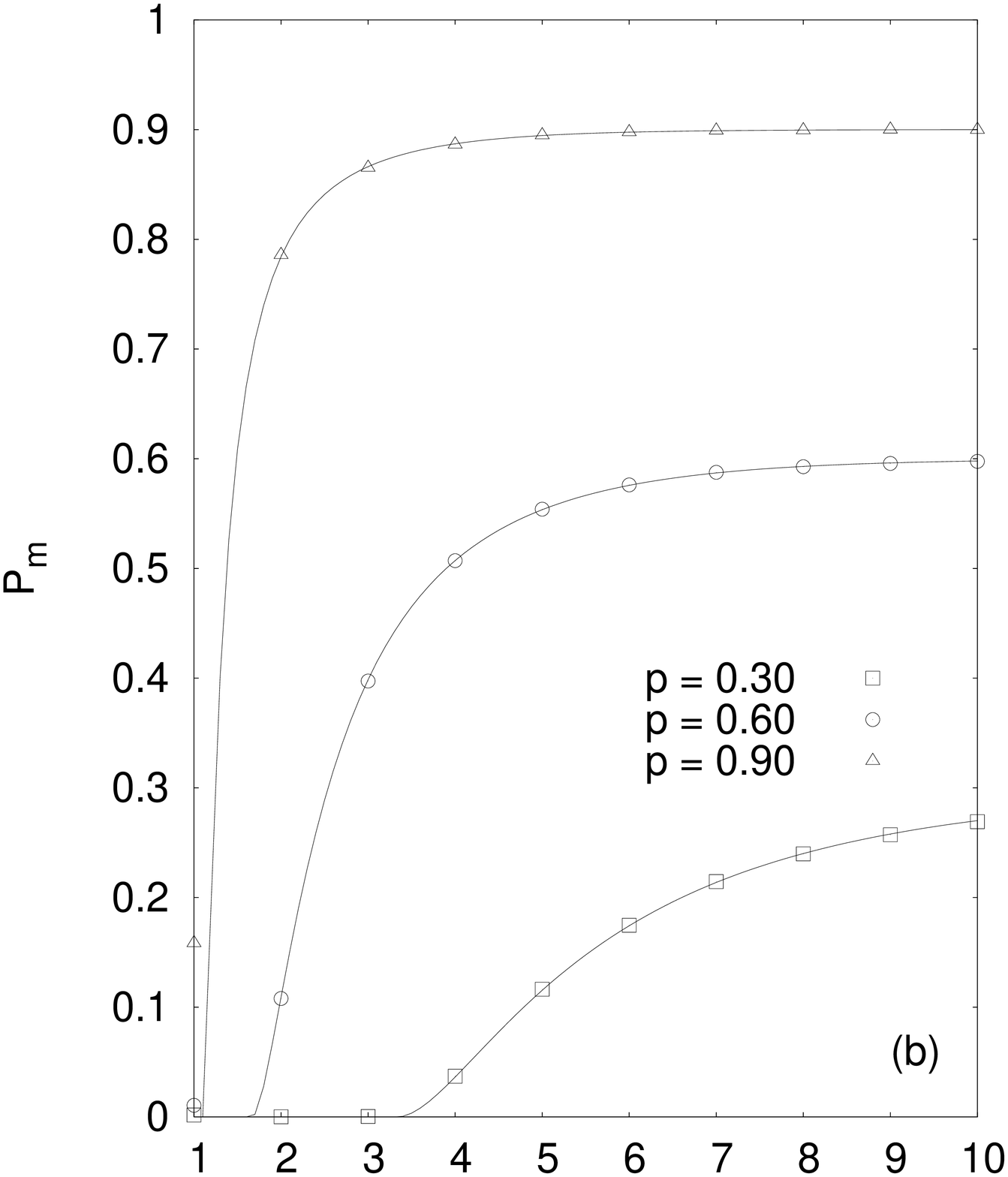}} &
   \resizebox*{\graphWidth}{!}{\includegraphics{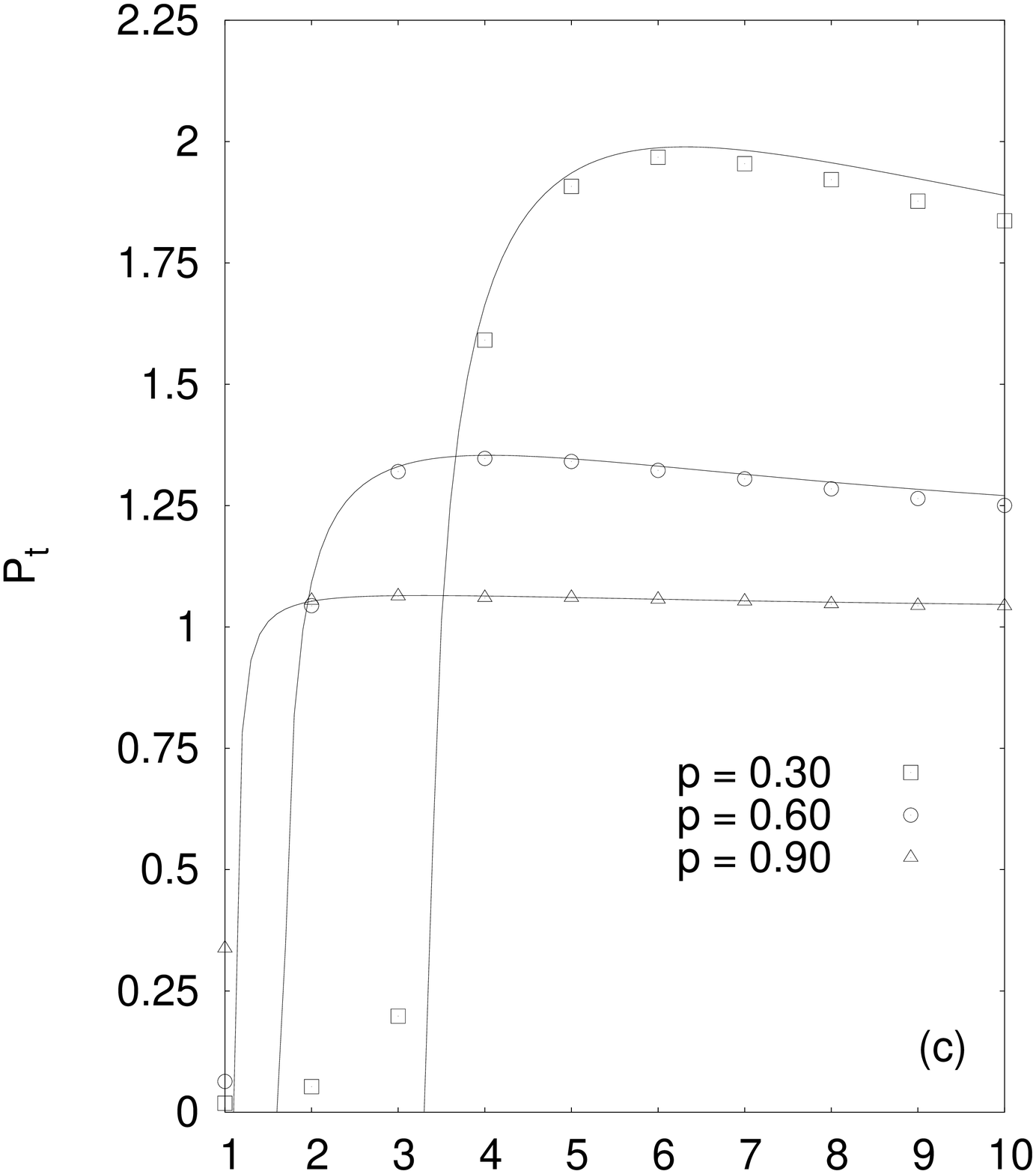}} \\
   \resizebox*{\graphWidth}{!}{\includegraphics{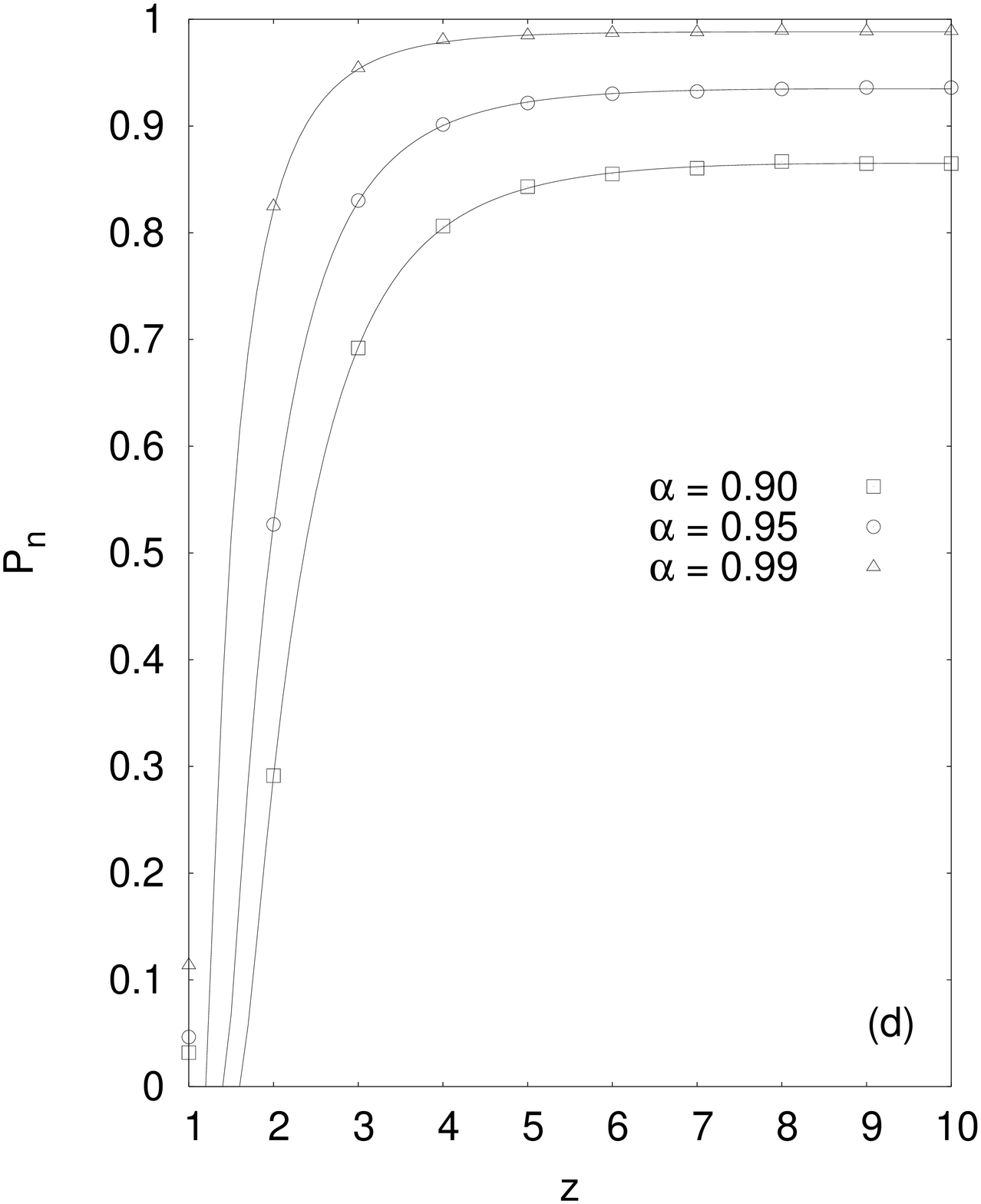}} &
   \resizebox*{\graphWidth}{!}{\includegraphics{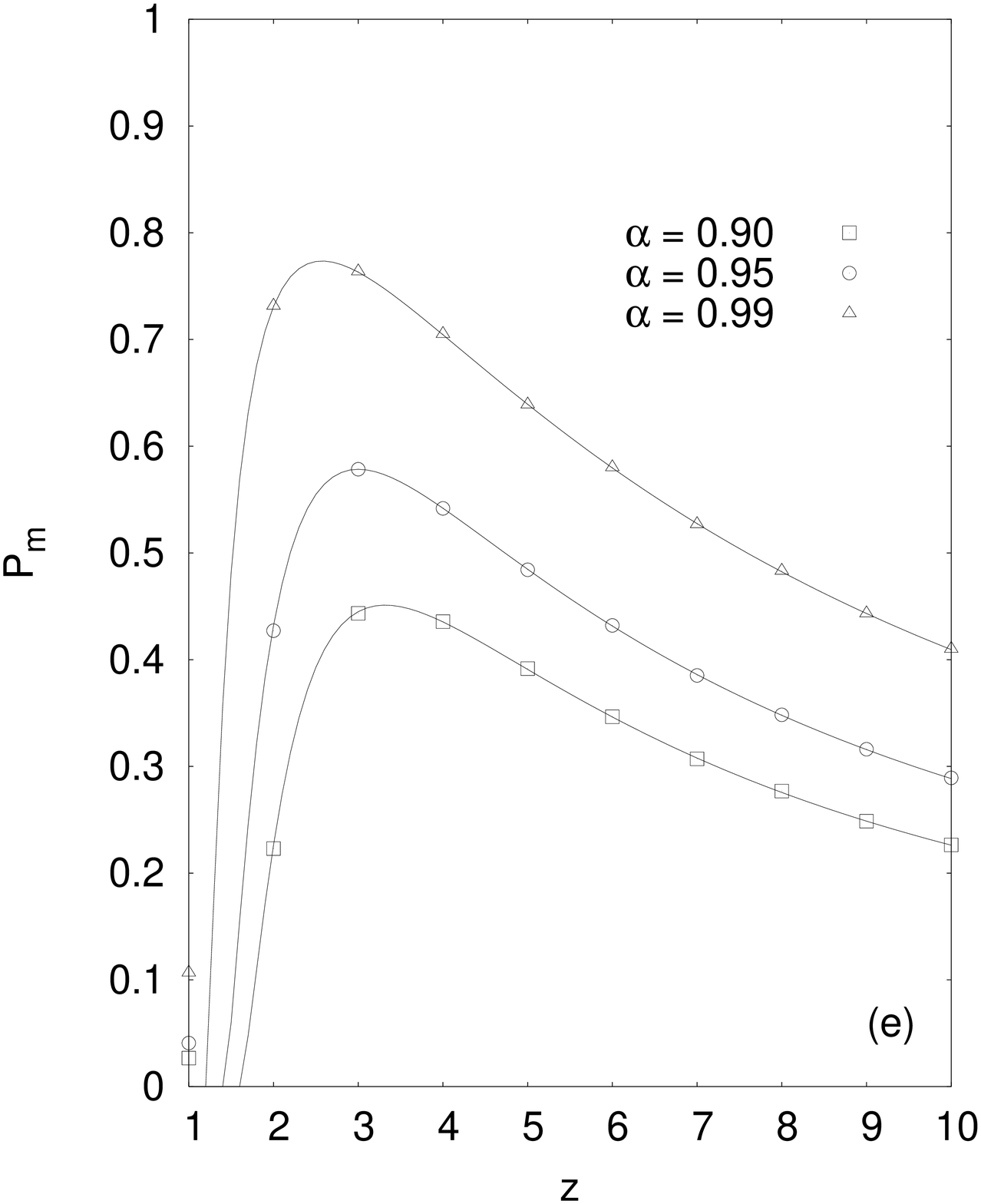}} &
   \resizebox*{\graphWidth}{!}{\includegraphics{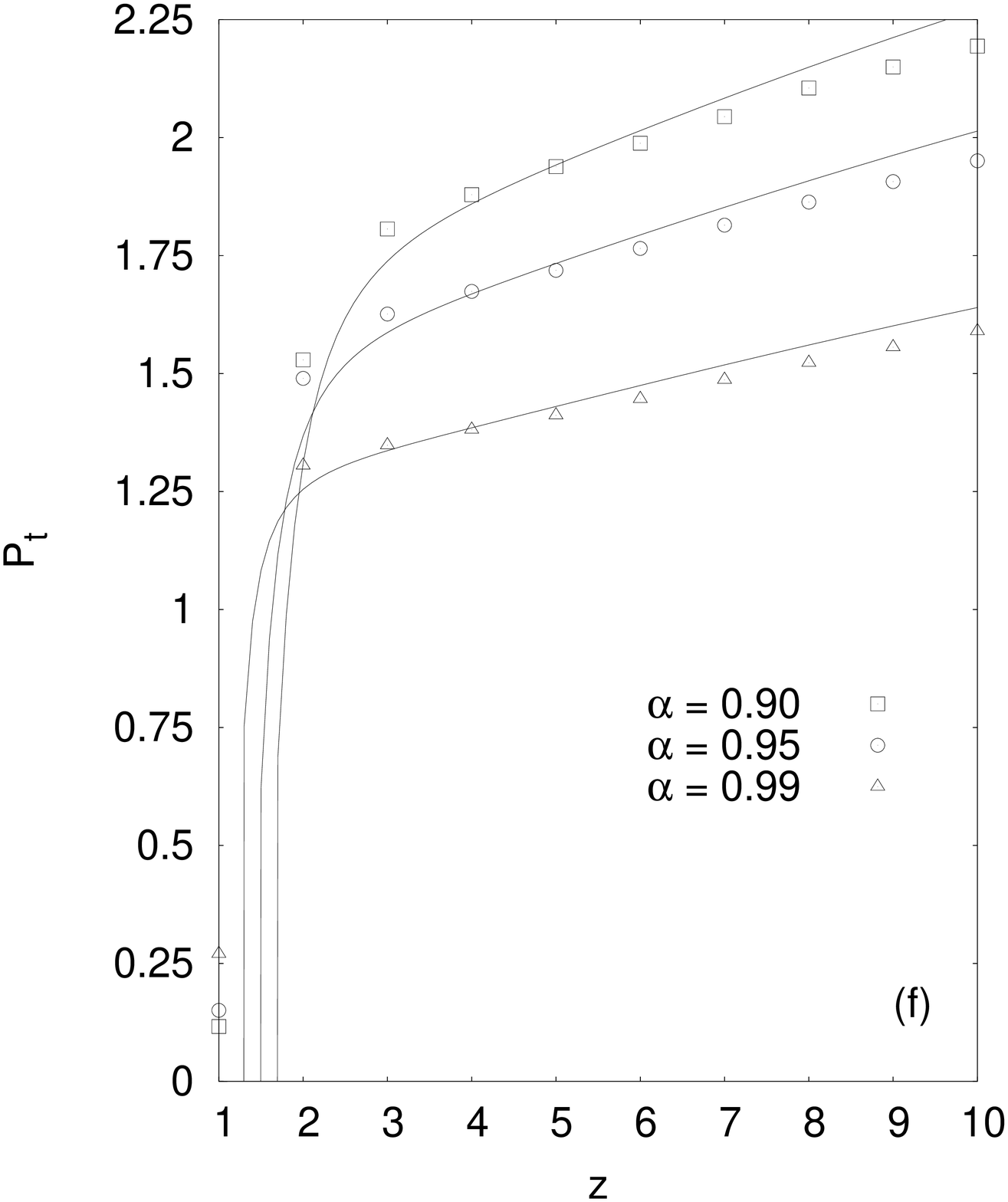}}
   \end{tabular}

   \caption{Simulation of probabilistic (a--c) and heuristic (d--f) flooding on random graphs with Poisson-distributed degrees.
   The plots show $\Pn$ (a and d), $\Pm$ (b and e), and $\Pt$ (c and f) for $p=0.30$, $0.60$, $0.90$ and $\alpha=0.90$, $0.95$, $0.99$.
   Solid lines give the analytical predictions of Section~\ref{sec:floodings}.
   }
   \label{fig:sim_poisson}
\end{figure*}

Figure~\ref{fig:sim_poisson} shows the results obtained by simulating probabilistic
and heuristic flooding on random graphs with Poisson-distributed degrees for $n=10000$ and $z$ varying from $1$ to $10$.
For each value of $z$, the simulation consisted of generating $15$ random graphs and, for each one, doing $1000$ instances
of each type of flooding (uninformed, probabilistic, and heuristic), each from a randomly chosen node inside the GCC. The simulation results are
given as averages of $\Pn$, $\Pm$, and $\Pt$ over the 15000 samples. Notice first of all that agreement between simulation and
analytical results is very good throughout, with a slight deviation only in parts (c) and (f) of the figure. Such deviations
are attributed to the various approximations of Section~$\ref{sec:pt}$.
Note also that both probabilistic and heuristic flooding do indeed improve on uninformed flooding as far as the number of
messages used is concerned ($\Pm < 1$), but nearly always at the expense of larger waiting times ($\Pt > 1$).

The results for probabilistic flooding (Figure~\ref{fig:sim_poisson}(a--c)) show that, given a value of $p$, there exists a value of
$z$ above which nearly all the nodes of the network are reached by the flooding. If we require $\Pn \geq 0.99$, then the threshold values
for $z$ are $z = 5$ for $p=0.90$ and $z = 9$ for $p=0.60$. For $p=0.30$ the threshold is greater than $10$. The plots for $\Pm$
show that $\Pm$ increases with $z$ roughly until these thresholds are reached, remaining constant and approximately equal to
$p$ from there onward. The plots for $\Pt$ show a similar but not identical behavior: like $\Pm$, $\Pt$ often increases roughly
until the thresholds are reached; unlike $\Pm$, $\Pt$ starts to decrease slightly as $z$ is further increased onward. All the plots
show that $\Pn$, $\Pm$, and $\Pt$ have value $0$ when $z$ is near $1$. This happens because, although the network is above the phase
transition that gives rise to the GCC of $G$ with high probability for any $z > 1$, for $z$ sufficiently near $1$ the flooding digraph $F$
is still below the phase transition at which the GIN and the GOUT appear (since $F$ is a directed subgraph of $G$), so they do not yet exist.

In the case of heuristic flooding (Figure~\ref{fig:sim_poisson}(d--f)), the results show that $\Pn$ flattens out much earlier than
probabilistic flooding to a value that depends on $\alpha$. This value is about $0.99$ for $\alpha=0.99$, $0.94$ for $\alpha=0.95$, and
$0.86$ for $\alpha=0.90$. The value of $\Pm$, in turn, increases sharply until approximately the value at which the flattening of $\Pn$
occurs. Past this value, $\Pm$ is seen to decrease continually, which indicates the desirable property that, given a fixed $\Pn$,
a progressively smaller fraction of messages is needed as the average degree increases. In contrast,
$\Pt$ increases onward through the higher values of $z$. This occurs because heuristic flooding tends to send messages on relatively
longer paths, avoiding as it does the sending of messages between high-degree nodes.

The second random-graph model we have considered is the one in which degrees are distributed according to a power law. In such a random
graph, the probability that a node has degree $a$ is $P_G(a)=C a^{-\tau}$, where $C$ is a normalizing constant and $\tau > 0$ is a parameter.
Clearly, in the limit as $n \to \infty$ we have $C=1/\zeta(\tau)$, where $\zeta(x)$ is the
Riemann zeta function \cite{yan}, that is, $\zeta(x)=\sum_{y=1}^{\infty} y^{-x}$.
We also have $Z_G=\zeta(\tau-1)/\zeta(\tau)$ and $\mean{K_G^2}=\zeta(\tau-2)/\zeta(\tau)$, so
the criterion of (\ref{eq:critical}) for the appearance of the GCC can be seen to translate into
\begin{equation}
   \frac{\zeta(\tau-2)}{\zeta(\tau-1)} > 2.
\end{equation}
Solving this inequality for $\tau$ numerically reveals that the GCC almost surely exists when $\tau < 3.47$.

Generation of a random graph with degrees thus distributed
can be achieved in two phases. First, the degrees $a_1, a_2, \ldots, a_n$ of the $n$ nodes, constituting the graph's so-called
degree sequence, are sampled from the power-law distribution and $\sum_{i=1}^n a_i$ labeled balls are put inside an imaginary
urn.\footnote{If $\sum_{i=1}^n a_i$ turns out to be odd, the degree sampling is repeated until an even sum is obtained.}
For $1 \leq i \leq n$, label $i$ is given to exactly $a_i$ balls.
In the second phase, a pair of balls (labeled, say, $u$ and $v$) is picked from the urn at random and an edge $(u,v)$ is added to the
graph. This process is repeated until the urn becomes empty. This algorithm clearly generates a multigraph (a graph in which multiple
edges and self-loops are allowed to exist), but the resulting graph is consistent with the model used in our analysis, where we assumed
edge independence throughout. Furthermore, even though there are other algorithms
to generate random graphs with a given degree sequence \cite{mihail2,mihail}, it is still unclear how to use them within reasonable
time bounds. The fact that multiple edges or self-loops may now exist in $G$ has direct impact on how the various flooding algorithms are 
simulated. Specifically, a node no longer considers whether to forward the information it receives for the first time to each of 
its neighbors, but rather whether to forward it on each of the edges that are incident to it in $G$.

\begin{figure*}[!t]
   \centering
   \begin{tabular}{ccc}
   \resizebox*{\graphWidth}{!}{\includegraphics{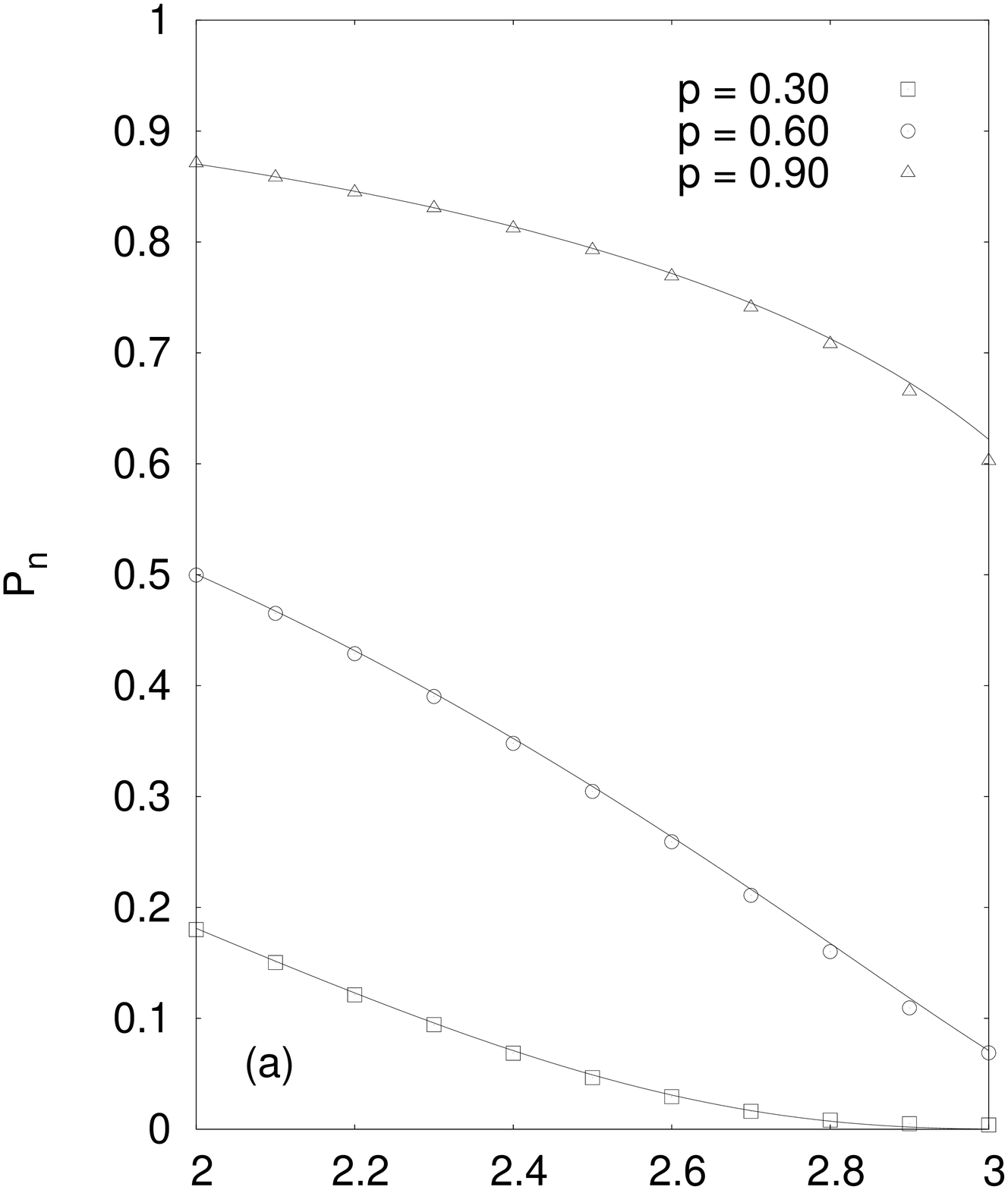}} &
   \resizebox*{\graphWidth}{!}{\includegraphics{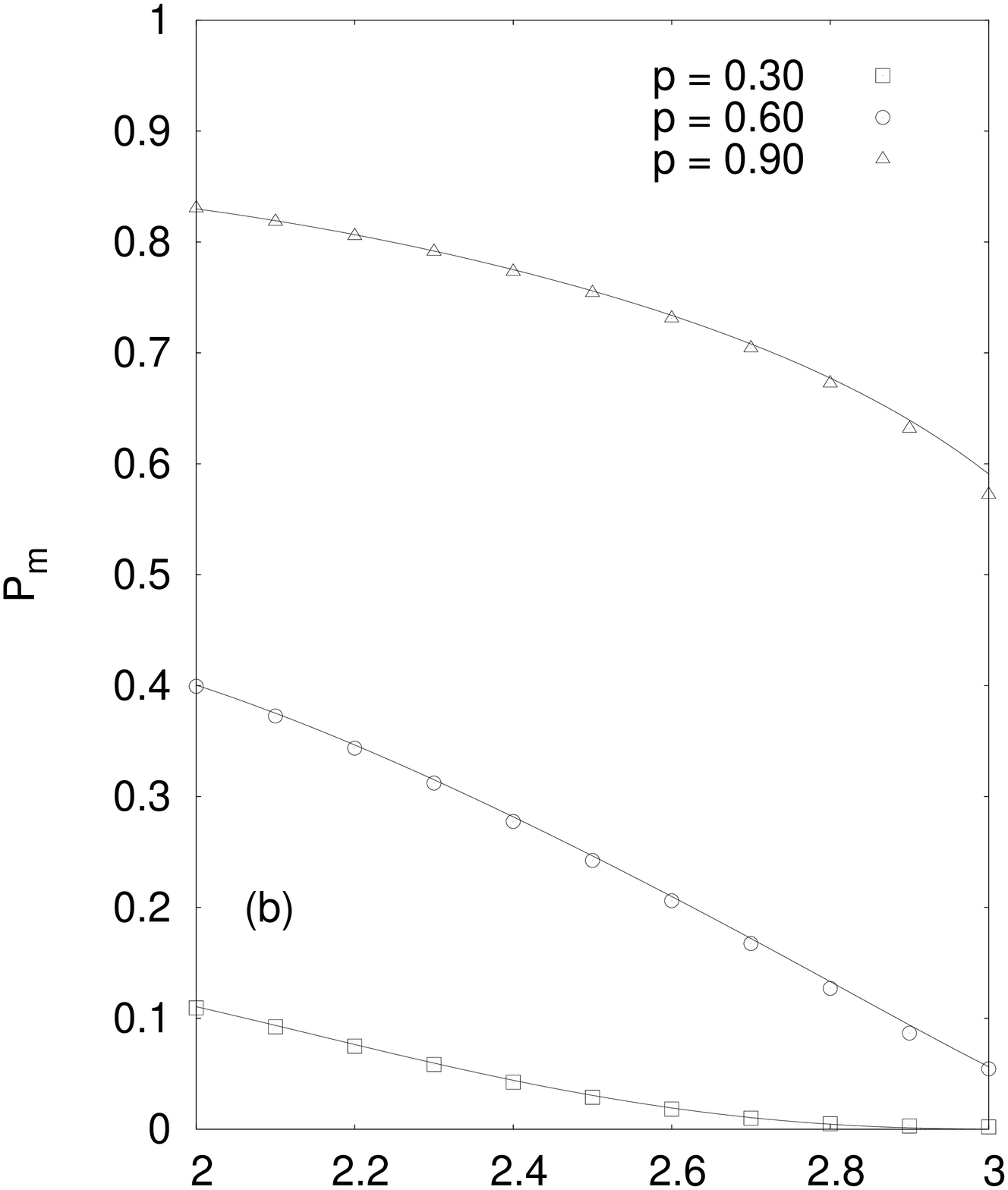}} &
   \resizebox*{\graphWidth}{!}{\includegraphics{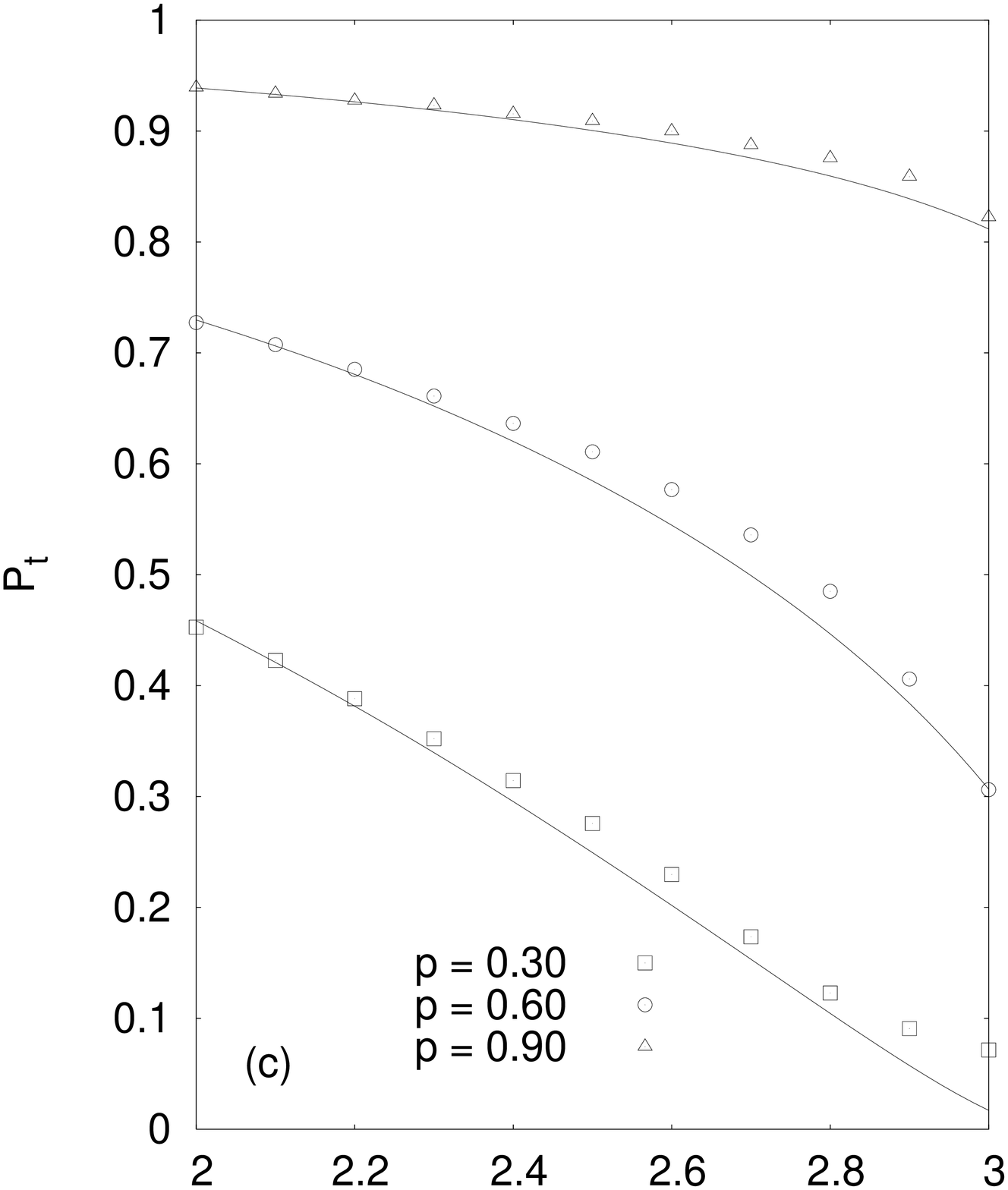}}\\
   \resizebox*{\graphWidth}{!}{\includegraphics{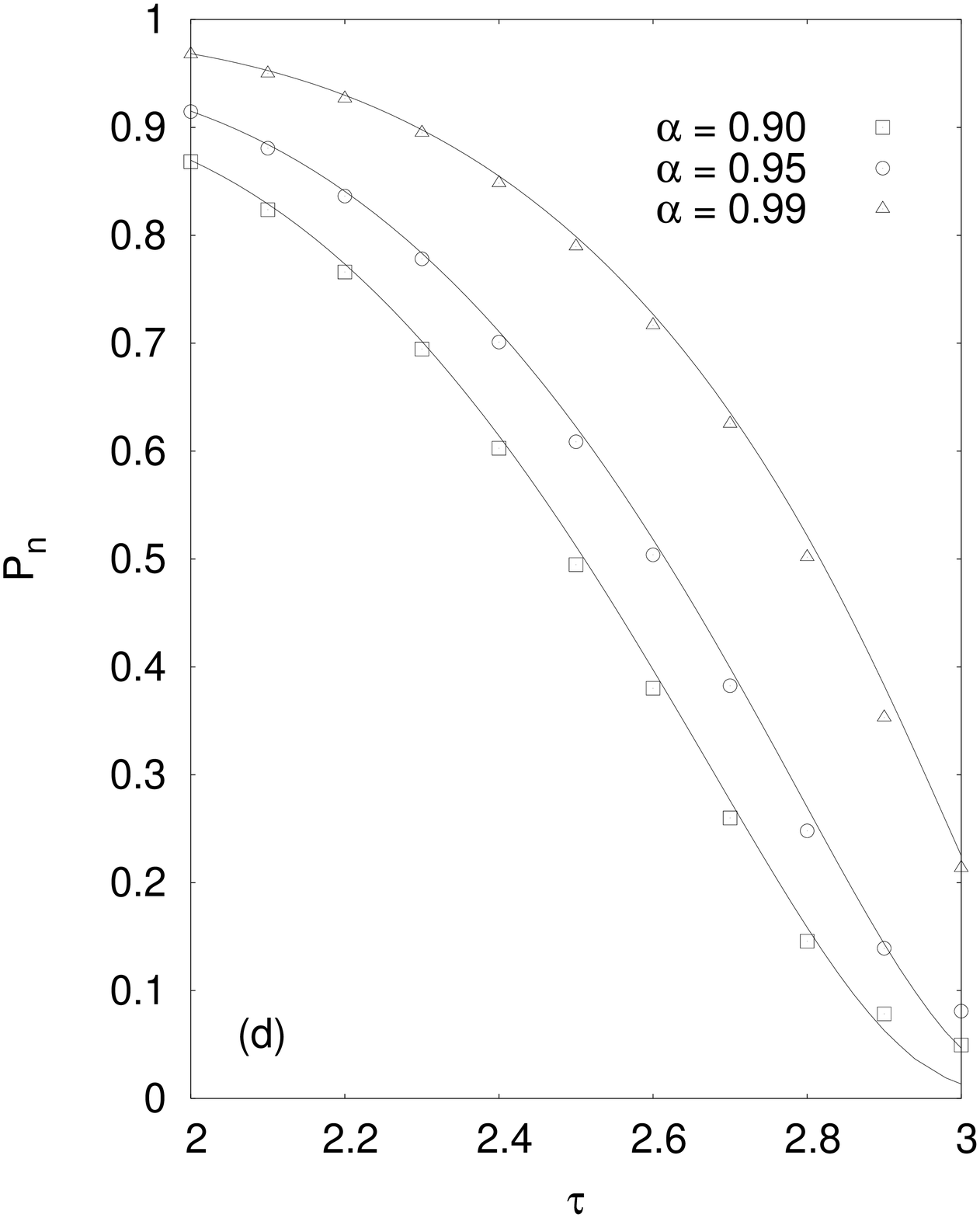}} &
   \resizebox*{\graphWidth}{!}{\includegraphics{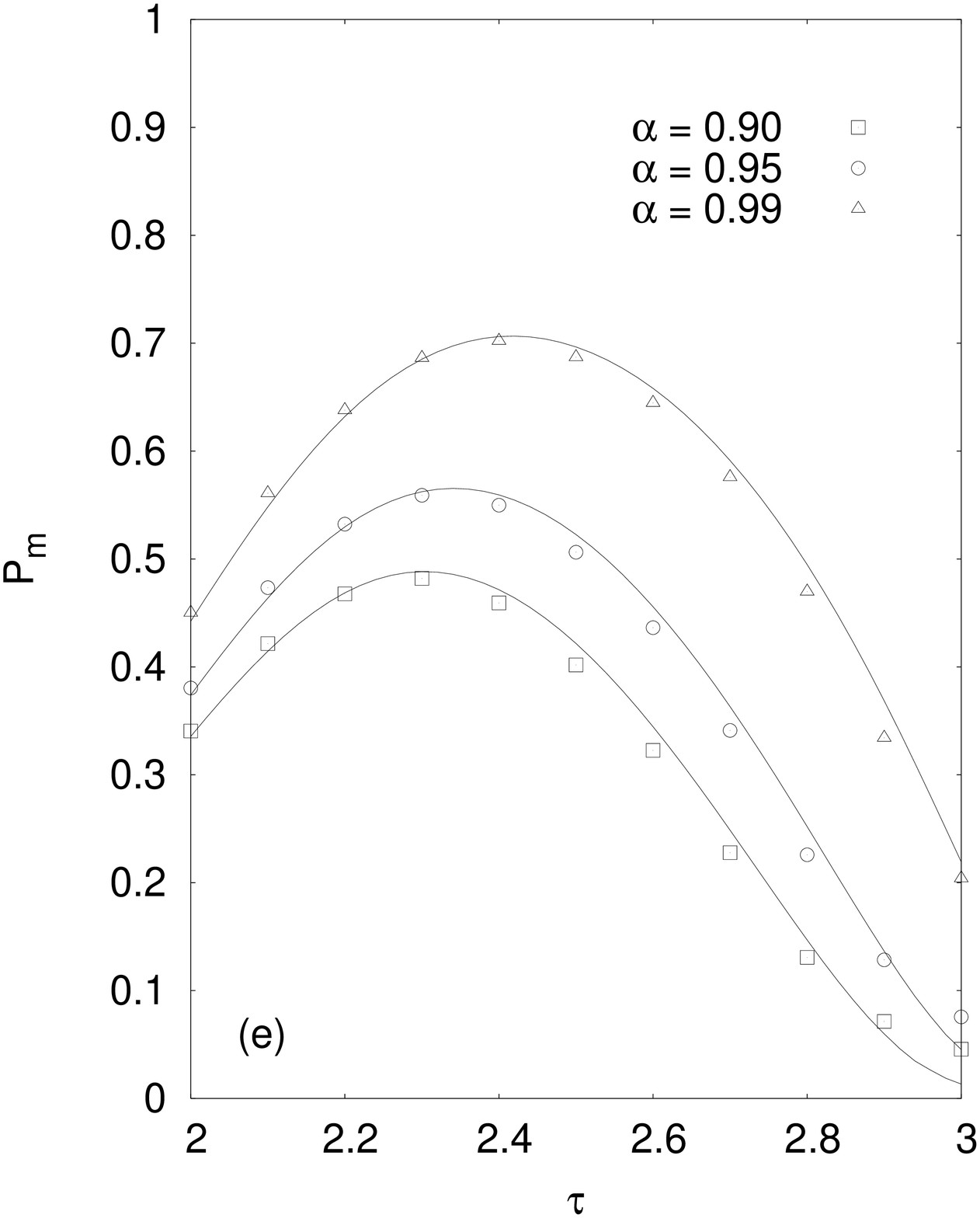}} &
   \resizebox*{\graphWidth}{!}{\includegraphics{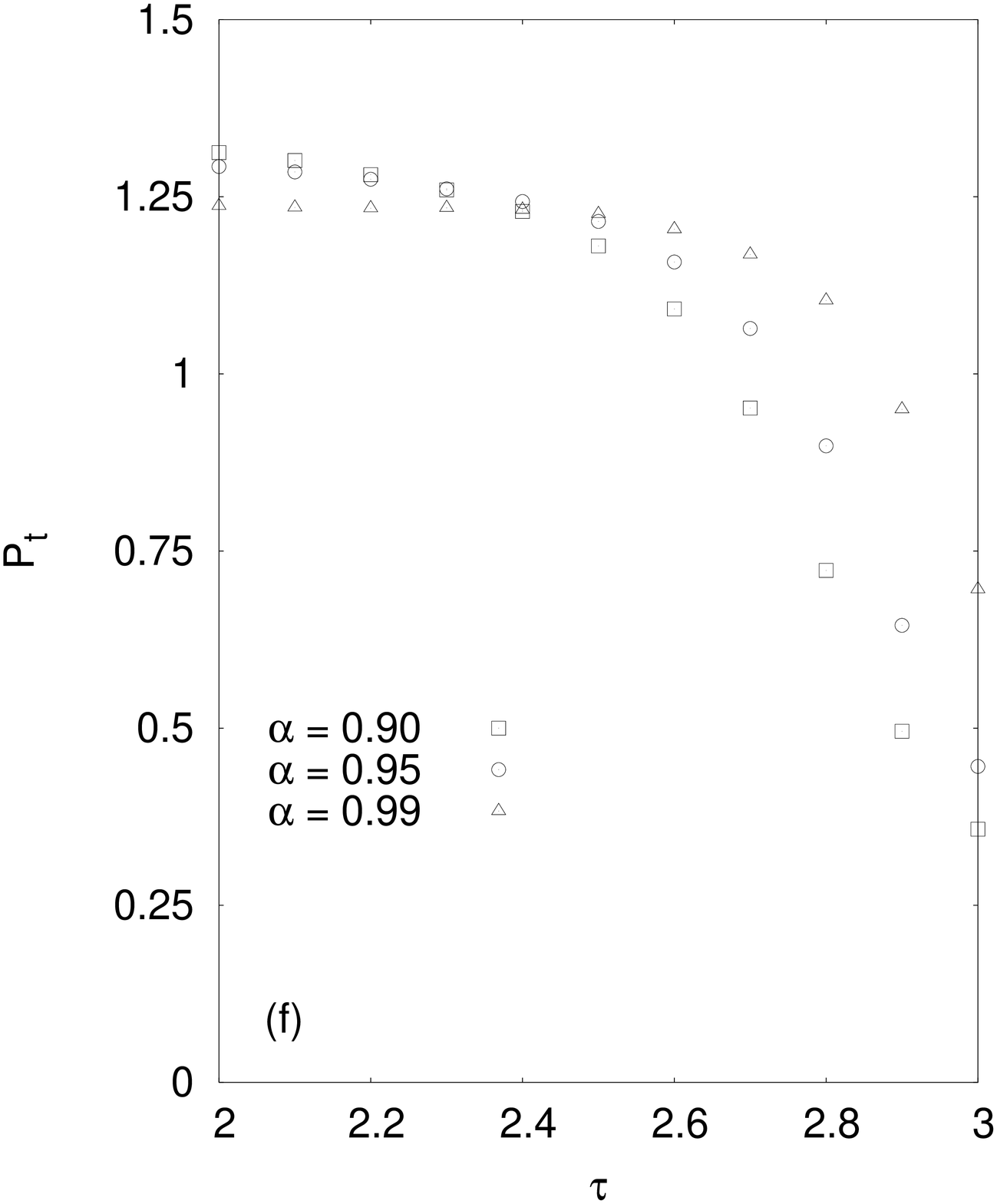}}
   \end{tabular}

   \caption{Simulation of probabilistic (a--c) and heuristic (d--f) flooding on random graphs with power-law-distributed degrees.
   The plots show $\Pn$ (a and d), $\Pm$ (b and e), and $\Pt$ (c and f) for $p=0.30$, $0.60$, $0.90$ and $\alpha=0.90$, $0.95$, $0.99$.
   Solid lines give the analytical predictions of Section~\ref{sec:floodings}.
   }
   \label{fig:sim_powerlaw}
\end{figure*}

Figure~\ref{fig:sim_powerlaw} shows the results obtained by simulating the three flooding methods
on random graphs with power-law-distributed degrees for $n=10000$ and $\tau$ varying from $2$ to $3$. For each
value of $\tau$, the simulation consists of generating $300$ random graphs and, for each one, proceeding exactly as indicated for the Poisson
case. First notice, again as in the Poisson case, that agreement between simulation and analytical results is very good for $\Pn$ and
$\Pm$. For $\Pt$, however, the situation is different. Under a power-law distribution for node degrees, expressions like (\ref{eq:z2gcc})
and (\ref{eq:z2gout}), respectively for $\ZGCC^{(2)}$ and $\ZGOUTr{2}$, are known to be problematic \cite{newman}, and indeed the
calculations do not converge and lead to wrong values for $\LGCC$ and $\LGOUT$. Even so, it is curious to note that for
probabilistic flooding the two errors seem to compensate each other somehow and the prediction for $\Pt$ comes close to the
simulation results, as shown in part (c) of the figure. The case of heuristic flooding, on the other hand, remains lacking a satisfactory
analytical prediction (thence none is shown in part (f) of the figure).

Also noteworthy in Figure~\ref{fig:sim_powerlaw} is the consistent superiority of probabilistic and heuristic flooding over
uninformed flooding regarding the number of messages sent ($\Pm < 1$). In addition, and contrasting with the Poisson case of
Figure~\ref{fig:sim_poisson}, the same holds for probabilistic flooding, and for sufficiently large $\tau$ also for
heuristic flooding, regarding waiting times ($\Pt < 1$). The price for this, of course, is heavy and reflected on
the often poor values of $\Pn$, with the exception of probabilistic flooding with $p=0.90$ and of heuristic flooding, more or less 
regardless of the value of $\alpha$, when $\tau$ is near $2$.

Simulation results for probabilistic flooding (Figure~\ref{fig:sim_powerlaw}(a--c)) show that $\Pn$ increases when $\tau$ decreases,
but the variation is sometimes not really too pronounced. For example,
for $p=0.90$ $\Pn$ is about $0.60$ when $\tau=3.0$ and $0.87$ when $\tau=2.0$. Similar observations are true of $\Pm$ and $\Pt$.
Notice, in addition, that unlike the Poisson case there does not seem to exist any interval within the range considered for $\tau$ inside
which any of the three indicators becomes approximately constant.

In heuristic flooding (Figure~\ref{fig:sim_powerlaw}(d--f)), the value of $\Pn$ also increases when $\tau$ decreases,
but at a faster pace than in the case of probabilistic flooding. The values of $\Pm$ and $\Pt$ display two different kinds of behavior.
While to either side of $\tau \approx 2.4$ the value of $\Pm$ decreases as $\tau$ is moved farther away, $\Pt$ falls relatively sharply 
to the right of $\tau \approx 2.4$ but rises, albeit slowly and with an inversion with respect to the values of $\alpha$, to the left.
From Figure~\ref{fig:sim_powerlaw}(d), $\tau \approx 2.4$ appears to coincide with a significant change in the second derivative of
$\Pn$ as a function of $\tau$.

\section{Probabilistic versus heuristic flooding}  \label{sec:comparison}

In spite of having already obtained results on both types of flooding, we still cannot compare them to each other directly, as
the parameter $p$ of probabilistic flooding has no relation whatsoever to the parameter $\alpha$ of heuristic flooding.
In this section, our strategy to make such a comparison possible is to first simulate heuristic flooding with $\alpha=0.99$ and then
use the resulting value of $\Pn$ to obtain, via (\ref{eq:pn}), the value of $p$ for which probabilistic flooding is expected to reach
the same fraction of the graph's nodes.

\begin{figure*}[!t]
   \centering

   \begin{tabular}{ccc}
   \resizebox*{\graphWidth}{!}{\includegraphics{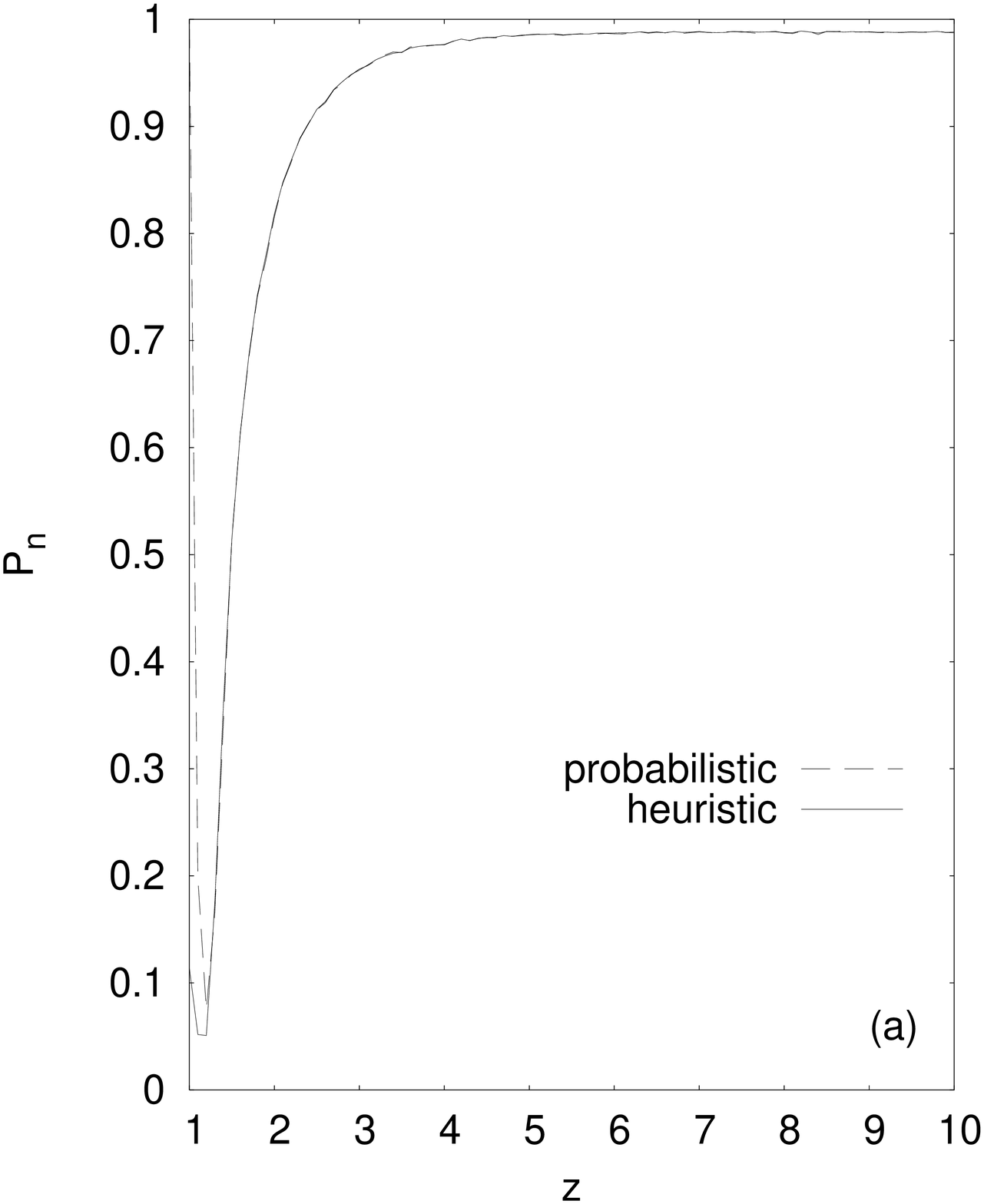}} &
   \resizebox*{\graphWidth}{!}{\includegraphics{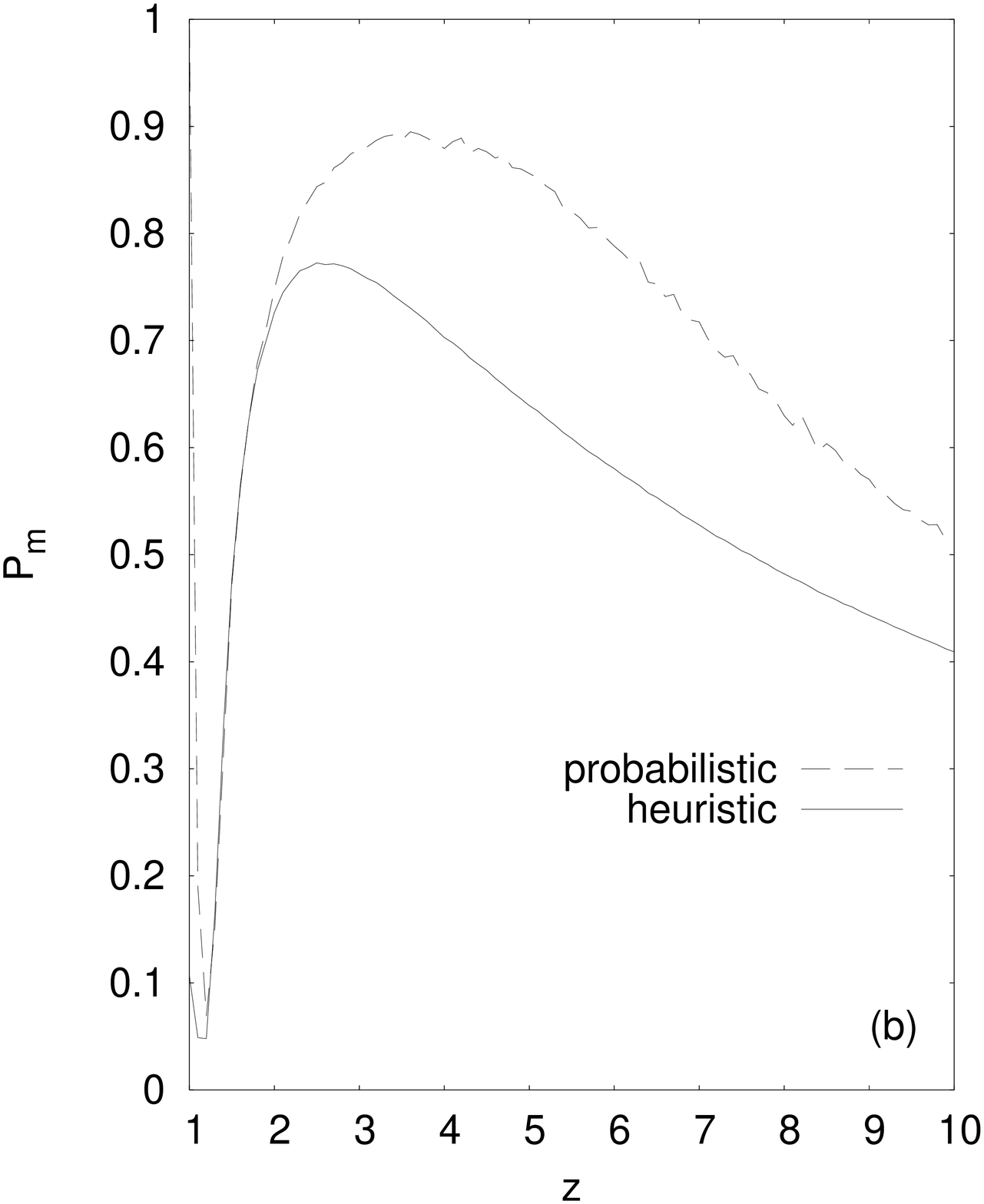}} &
   \resizebox*{\graphWidth}{!}{\includegraphics{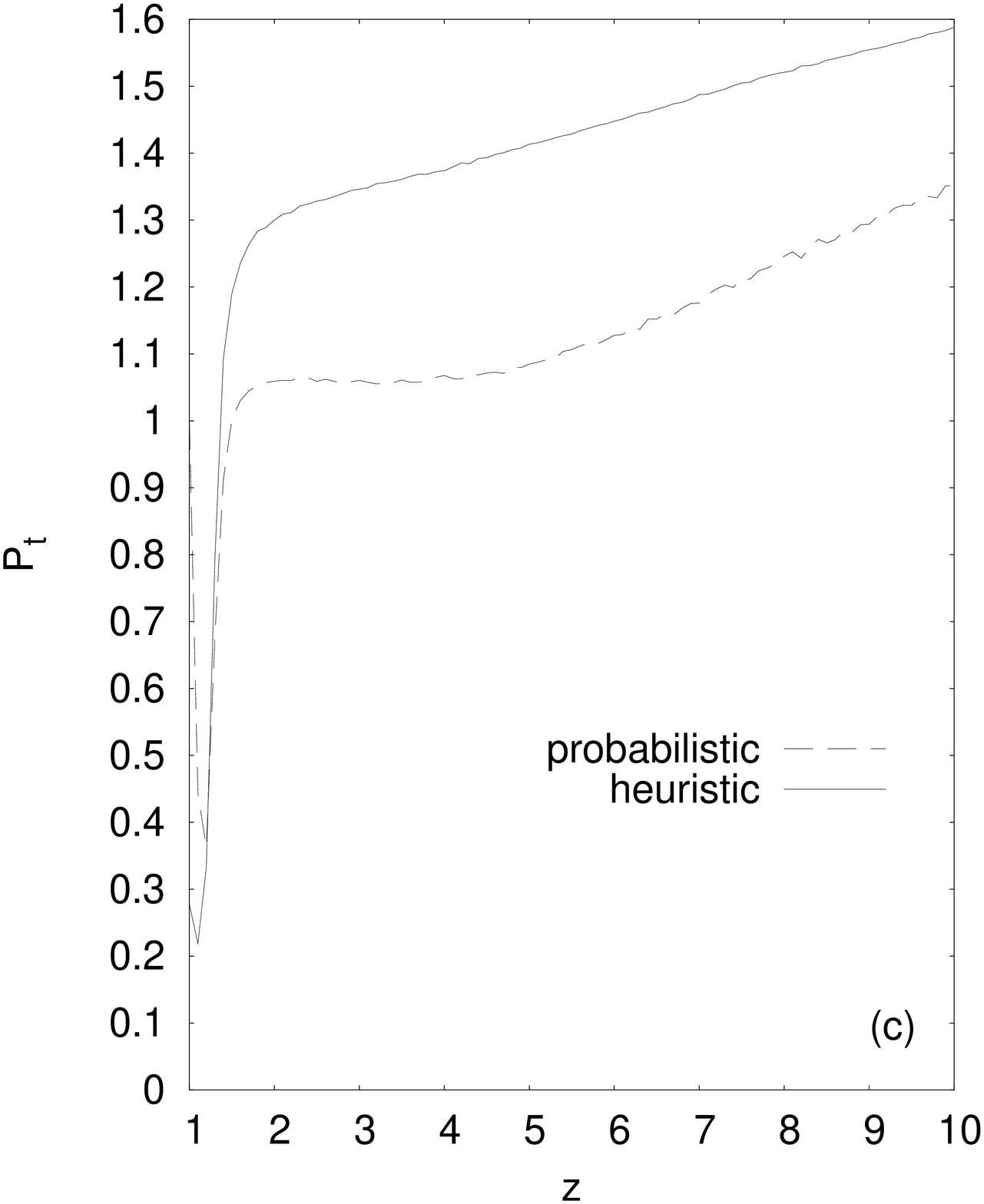}}
   \end{tabular}

   \caption{Comparison of probabilistic flooding and heuristic flooding in the Poisson case. Plots are shown for
      $\Pn$ (a), $\Pm$ (b), and $\Pt$ (c).
   }
   \label{fig:sim_poisson_cmp}
\end{figure*}

Our results under this strategy are shown in Figure~\ref{fig:sim_poisson_cmp} for the Poisson case. Notice, first, that the
strategy seems indeed to be effective, since the plots for $\Pn$ are practically the same for both types of flooding.
For $z$ right past $1$, all the six plots exhibit an anomaly, which is once again likely due to the fact that, at these values of
$z$, the phase transition that gives rise to the GIN and the GOUT has not yet taken place.

What is most interesting, though, is the evident trade-off between $\Pm$ and $\Pt$ that can be seen in parts (b) and (c) of the
figure. Clearly, under the constraint that $\Pn$ is roughly the same for both probabilistic and heuristic flooding, for $z$ larger
than approximately $2.5$, $\Pm$ is significantly larger for probabilistic flooding than it is for heuristic flooding. On the other
hand, $\Pt$ is significantly larger for heuristic flooding across practically all the spectrum of $z$ values that Figure
\ref{fig:sim_poisson_cmp} encompasses.

But notwithstanding this trade-off, Figure~\ref{fig:sim_poisson_cmp} also suggests that the two types of flooding become
more and more similar to each other as $z$ is increased. In fact, as the nodes' degrees become larger and $G$ more homogeneous
(approaching, in the limit, the complete graph on $n$ nodes), the heuristic function of (\ref{eq:heuristic}) approaches a small constant
and heuristic flooding progressively becomes probabilistic flooding.

\begin{figure*}[!t]
   \centering

   \begin{tabular}{ccc}
   \resizebox*{\graphWidth}{!}{\includegraphics{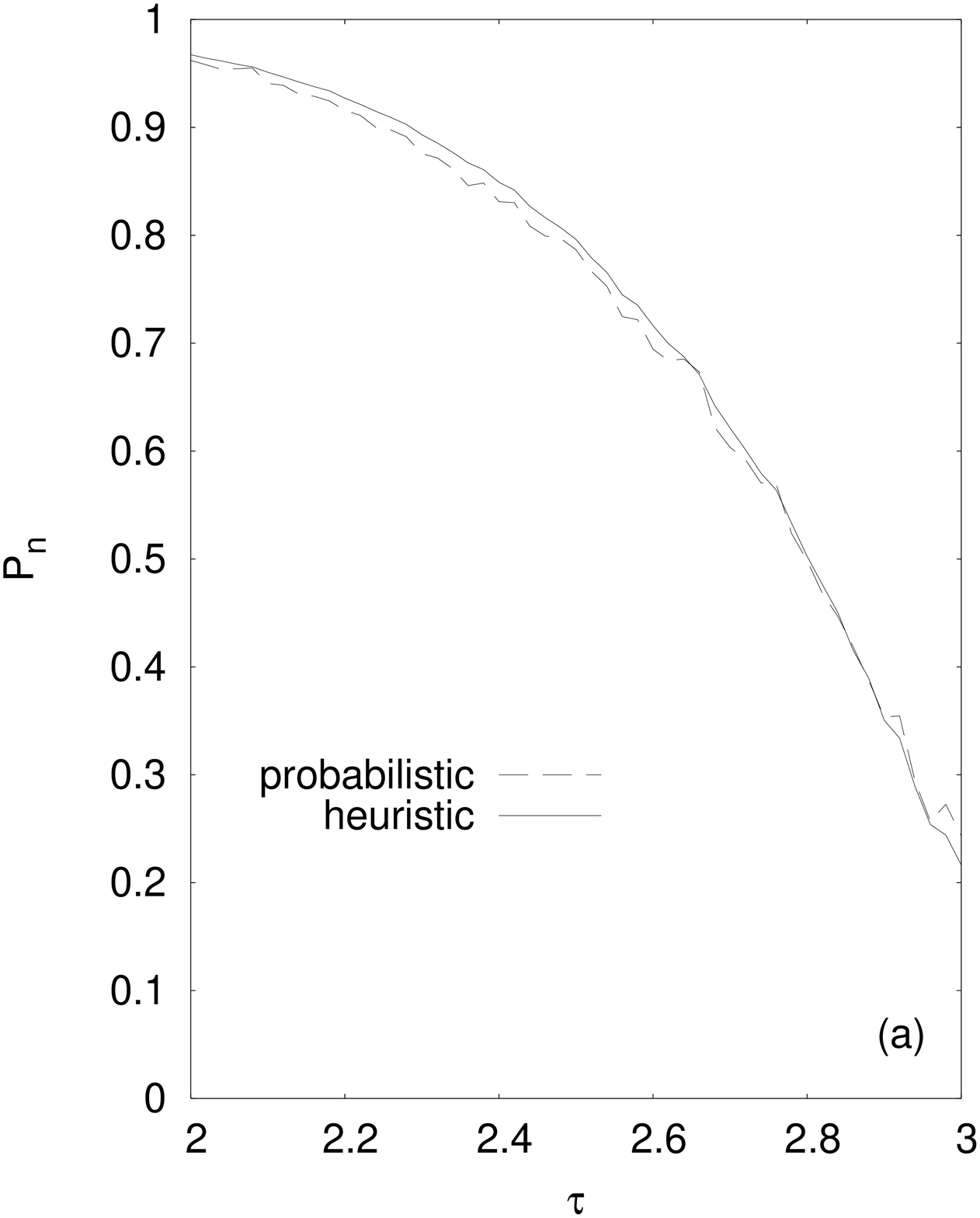}} &
   \resizebox*{\graphWidth}{!}{\includegraphics{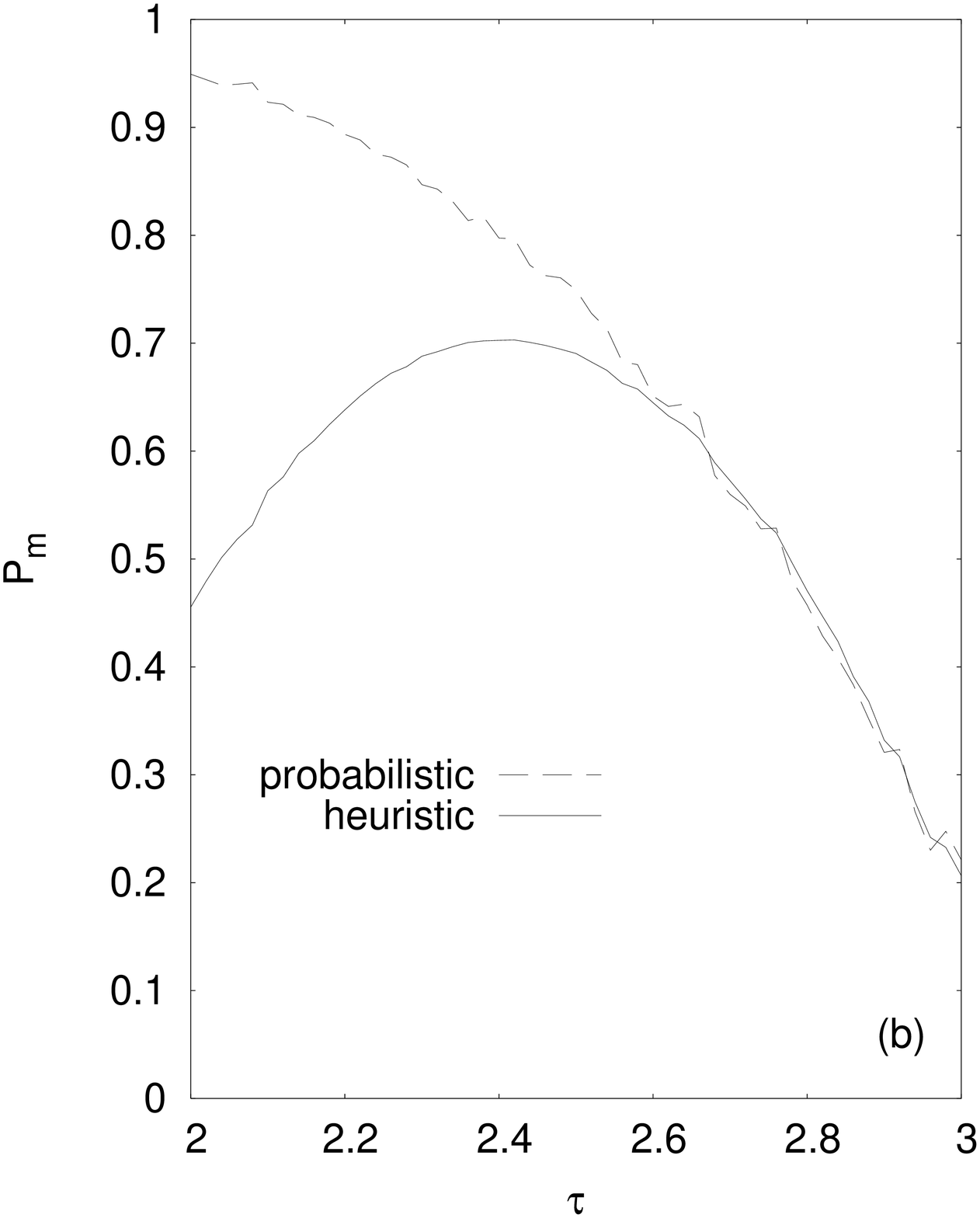}} &
   \resizebox*{\graphWidth}{!}{\includegraphics{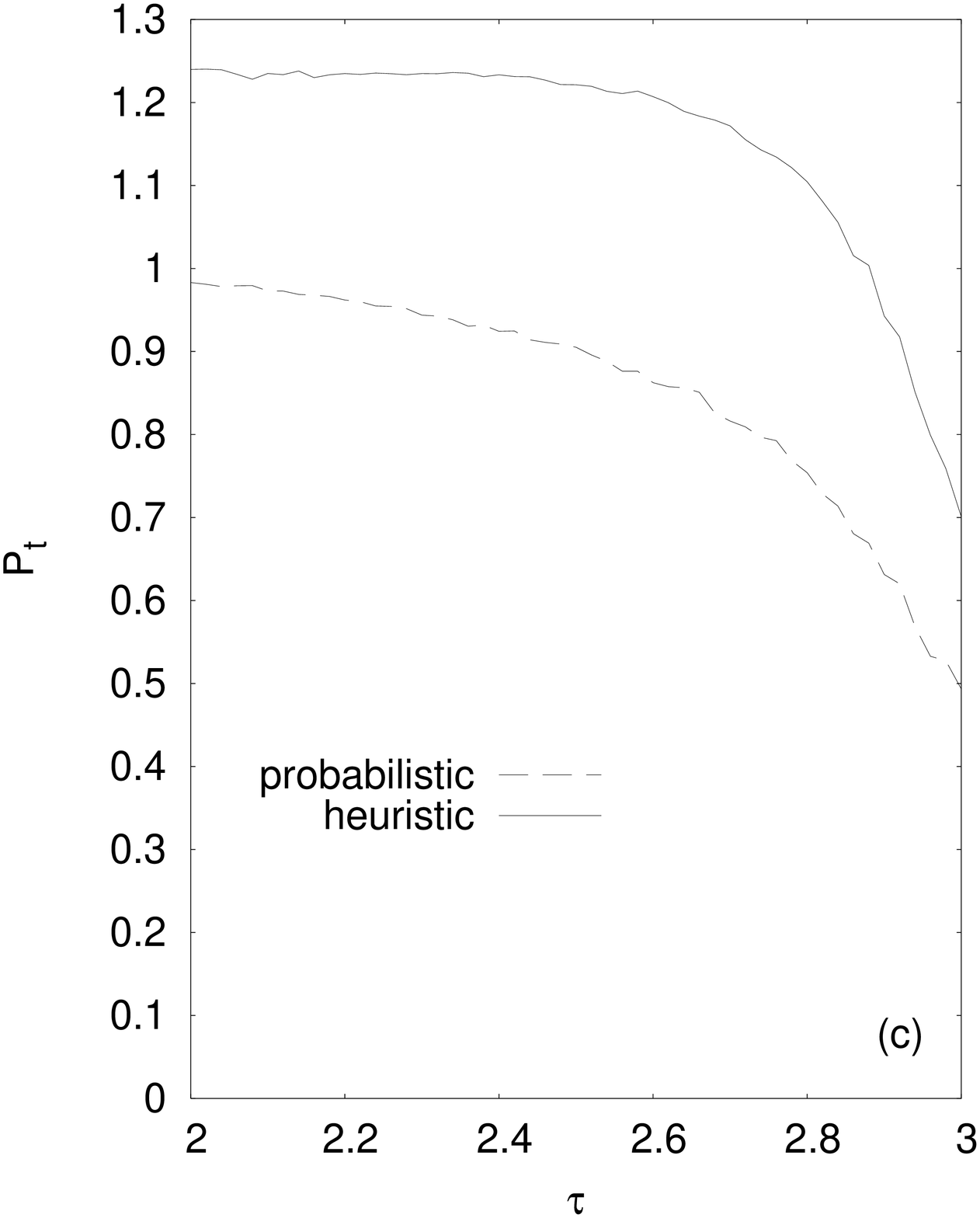}}
   \end{tabular}

   \caption{Comparison of probabilistic flooding and heuristic flooding for the power-law case. Plots are shown for
      $\Pn$ (a), $\Pm$ (b), and $\Pt$ (c).
   }
   \label{fig:sim_powerlaw_cmp}
\end{figure*}

The corresponding results for graphs with degrees distributed according to a power law are shown in the plots of Figure
\ref{fig:sim_powerlaw_cmp}. Once again, our strategy's effectiveness is corroborated by part (a) of the figure. Also,
parts (b) and (c) reveal the same trade-off between $\Pm$ and $\Pt$ that we observed in the Poisson case, along with a similar
tendency, especially in the case of part (b), for the two algorithms to resemble each other as $\tau$ is increased toward $3$.
The justification for part (b) is that, at these values for $\tau$, heuristic flooding has a small value for $\Pn$ (cf.~part (a)), one
that can be achieved by probabilistic flooding with a value of $p$ for which $\Pm$ remains relatively low. As for part (c), increasing
$\tau$ toward its higher values makes the occurrence of nodes of very high degree ever less likely. Consequently, heuristic flooding
no longer refrains so strongly from sending the information being broadcast to the high-degree nodes of the network. This makes path
lengths tend to lean toward those of probabilistic flooding.

But what is especially noteworthy in the power-law case is that the superiority of heuristic flooding in terms of $\Pm$
becomes very pronounced as $\tau$ is decreased from roughly $\tau=2.5$. This happens
because, under a power law, the graph tends to contain a large set of nodes of small degree, many of them of degree $1$. In
order to reach the same number of nodes as heuristic flooding, probabilistic flooding must run with a value for $p$ that is
near $1$, which causes an unnecessarily high number of messages to be sent. Heuristic flooding, in turn, is relatively
insensitive to the plethora of low-degree nodes, since it attempts to provide each and every node with the same probability
of receiving the information.

\section{Conclusions}  \label{sec:conclusion}

Flooding a network probabilistically for information dissemination is an attempt at reducing the heavy communications demand that
uninformed flooding incurs in terms of how many messages are needed. The drawback, of course, is that the guarantee of network-wide
information delivery may be lost. In this paper we started with the recently introduced probabilistic flooding, in which every
node forwards the information to all of its neighbors with constant probability.

Then we introduced an alternative technique, called heuristic flooding, which employs degree-based probabilistic decisions at the nodes,
aiming at a pre-defined probability, the same for all nodes, that a node receives the information being disseminated. We have also
contributed a detailed mathematical analysis of both probabilistic flooding and heuristic flooding, comprising analytical predictions
of the techniques' reachability, their communications requirements, and an indicator related to the time needed for completion
of the flooding.

Our extensive simulation results have revealed the two techniques' main characteristics on random graphs with Poisson- and 
power-law-distributed degrees. They have also indicated an excellent agreement between theory and experimentation in most cases.

In addition, one final set of simulations designed especially to allow a meaningful comparison between the two flooding techniques
demonstrated an interesting trade-off involving them: in general, heuristic flooding outperforms probabilistic flooding in terms of
communications requirements, but the opposite holds in terms of the delay required for completion. Curiously, the balance between the
two sides of this trade-off is not equally tipped: for example, under a power law with relatively small parameter, the
communications-related gain of heuristic flooding over probabilistic flooding significantly surpasses its delay-related loss.

\subsection*{Acknowledgments}
The authors acknowledge partial support from CNPq, CAPES, and a FAPERJ BBP grant.

\bibliography{heuristicFlooding}
\bibliographystyle{plain}

\end{document}